\documentclass[12pt]{article}
\usepackage{epsfig}

\usepackage{graphicx}
\usepackage{cancel}
\usepackage{amssymb,amsmath}

\def\harr#1#2{\smash{\mathop{\hbox to .3in{\rightarrowfill}}
 \limits^{\scriptstyle#1}_{\scriptstyle#2}}}

\def\s2{\frac{1}{\sqrt2}}

\def\be{\begin{equation}}
\def\ee{\end{equation}}
\def\beqa{\begin{eqnarray}}
\def\eeqa{\end{eqnarray}}

\def\Dsl{\,\raise.15ex\hbox{/}\mkern-13.5mu D} 
\def\d3{d^3}

\begin{document}

\begin{center}
\Large{\bf On $p$-adic string amplitudes in the limit $p$ approaches
to one}
\vspace{0.5cm}

\large  M. Bocardo-Gaspar$^a$\footnote{e-mail address: {\tt
mbocardo@math.cinvestav.mx}},
           H. Garc\'{\i}a-Compe\'an$^{b}$\footnote{e-mail address: {\tt compean@fis.cinvestav.mx}},
           W. A. Z\'{u}\~{n}iga-Galindo$^{a}$\footnote{e-mail address: {\tt wazuniga@math.cinvestav.edu.mx}}\\

\vspace{0.5cm}

{\small \em Centro de Investigaci\'{o}n y de Estudios Avanzados
del Instituto Polit\'{e}cnico Nacional\\
$^a$Departamento de Matem\'{a}ticas, Unidad Quer\'{e}taro\\
Libramiento Norponiente \#2000, \\
Fracc. Real de Juriquilla. Santiago de Quer\'{e}taro, Qro. 76230,
M\'{e}xico}

{\small \em $^b$Departamento de F\'{\i}sica,}\\
{\small\em P.O. Box 14-740, CP. 07000, M\'exico D.F., M\'exico.}\\

\vspace*{0.5cm}
\end{center}

\begin{abstract}
In this article we discuss the limit $p$ approaches to one of
tree-level $p$-adic open string amplitudes and its connections with
the topological zeta functions. There is empirical evidence that
$p$-adic strings are related to the ordinary strings in the $p \to
1$ limit. Previously, we established that $p$-adic Koba-Nielsen
string amplitudes are finite sums of multivariate Igusa's local zeta
functions, consequently, they are convergent integrals that admit
meromorphic continuations as rational functions. The meromorphic
continuation of local zeta functions has been used for several
authors to regularize parametric Feynman amplitudes in field and
string theories. Denef and Loeser established that the limit $p \to
1$ of a Igusa's local zeta function gives rise to an object called
topological zeta function. By using Denef-Loeser's theory of
topological zeta functions, we show that limit $p \to 1$ of
tree-level $p$-adic string amplitudes give rise to certain
amplitudes, that we have named Denef-Loeser string amplitudes.
Gerasimov and Shatashvili showed that in limit $p \to 1$ the
well-known non-local effective Lagrangian (reproducing the
tree-level $p$-adic string amplitudes) gives rise to a simple
Lagrangian with a logarithmic potential. We show that the Feynman
amplitudes of this last Lagrangian are precisely the amplitudes
introduced here. Finally, the amplitudes for four and five points
are computed explicitly.

\vskip .3truecm

\noindent\leftline{May 04, 2018}
\end{abstract}

\bigskip
\newpage
\section{Introduction}
\label{sec:intro}

The $p$-adic field and string theories have been studied over the
time with some periodic fluctuations in their interest (for some
reviews, see
\cite{Hlousek:1988vu,Brekke:1993gf,V-V-Z,Dragovich:2017kge}).
Recently a considerable amount of work has been performed on this
topic in the context of the AdS/CFT correspondence
\cite{Gubser:2016guj,Heydeman:2016ldy,Gubser:2016htz,Dutta:2017bja}.

On the other hand, Sen's conjecture asserts that the tachyonic
potential has a local minimum which exactly cancels the total energy
of a D-brane in string theory \cite{Sen:1998sm,Sen:1999xm}. This
conjecture has been proved by using bosonic and superstring field
theories \cite{Sen:1999nx,Berkovits:2000hf}. In the $p$-adic
setting, Sen's conjecture is easier to verify than the classical
version, see \cite{Ghoshal:2000dd}.

In string theory, $N$-point string amplitudes are an important
observable, which is computed through integration over the moduli
space of Riemann surfaces \cite{Berera:1992tm}. It is known that
even at the tree-level amplitudes the convergence of these integrals
have not been well understood for general $N$ \cite{Witten:2013pra}.
For particular values of $N$, for instance $N=4$ or $N=5$ for open
and closed strings at the tree-level there are some criteria for an
appropriate choice of the external momenta in such a way that the
corresponding integrals converge and the corresponding amplitudes
are well defined.

In \cite{Brekke:1988dg} (see also \cite{Frampton:1987sp}), an
effective Lagrangian was proposed from which there can be derived
the Feynman rules necessary to compute the $N$-point $p$-adic string
amplitudes at tree-level.  Later, some time-dependent solutions to
the effective action have been found representing a rolling tachyon
for potentials for both $p$ even and odd \cite{Moeller:2002vx}.
Moreover this effective action has been used also with cosmological
purposes, for instance inflation was studied in
\cite{Barnaby:2006hi}.

The $p$-adic strings seem to be related in some interesting ways
with ordinary strings. For instance, connections through the adelic
relations \cite{Freund:1987ck} and through the limit when $p \to 1$
\cite{Gerasimov:2000zp,Spokoiny:1988zk}, have been discussed in the
literature. In  \cite{Gerasimov:2000zp}, the limit $p\to 1$ of the
effective action was studied, it was showed that this limit gives
rise to a boundary string field theory (BSFT), which was previously
proposed by Witten in the context of background independent string
theory \cite{Witten:1992qy,Witten:1992cr}. The limit $p\to 1$ in the
effective theory can be performed without any problem. Though
originally $p$ was a prime number for the world-sheet theory, in the
effective theory one can consider $p$ just as integer or real
parameter and take formally the limit $p \to 1$. The resulting
theory is related to a field theory describing an open string
tachyon \cite{Minahan:2000ff}. In the limit $p \to 1$ also there are
exact noncommutative solitons, some of these solutions were found in
\cite{Ghoshal:2004dd}. Moreover, this limit has found a very
interesting physical interpretation in \cite{Ghoshal:2006te}, in
terms of a lattice discretization of ordinary string worldsheet. In
the worldsheet theory we cannot forget the nature of $p$ as a prime
number, thus the analysis of the limit is more subtle. The correct
way of taking the limit $p \to 1$ involves the introduction of
finite extensions of the $p$-adic field $\mathbb{Q}_p$. The totally
ramified extensions gives rise to a finer discretization of the
worldsheet following the rules of the renormalization group
\cite{Ghoshal:2006te}. In this article we will also require the use
of finite extensions of the $p$-adic field at the level of the
string amplitudes.

In \cite{Bocardo-Gaspar:2016zwx}, we showed that the $p$-adic open
string $N-$point tree amplitudes are bonafide integrals that admit
meromorphic continuations as rational functions, by relating them
with multivariate local zeta functions (also called multivariate
Igusa local zeta functions \cite{Igusa,Meuser}). Moreover Denef and
Loeser \cite{denefandloeser} established that the limit $p$
approaches to one of a local zeta function give rise a new object
called a {\it topological zeta function}, which is associated with a
complex polynomial. By using the theory of topological zeta
functions, we show that limit $p \to 1$ of $p$-adic string
amplitudes gives rise to certain string amplitudes, that we called
{\it Denef-Loeser open string amplitudes} which are rational
functions. Taking the limit at the level of Koba-Nielsen amplitudes
involves the introduction of finite extension of the $p$-adic field
$\mathbb{Q}_p$. This task is carried out here using the results of
\cite{Bocardo-Gaspar:2016zwx}.

Finally, we want to point out that the results presented in this
article are essentially independent of the results given in
\cite{Bocardo-Gaspar:2016zwx}. More precisely, in order to use
Denef-Loeser's theory of topological zeta functions, we do not need
the convergence of the $p$-adic Koba-Nielsen string amplitudes which
is one of the main results in \cite{Bocardo-Gaspar:2016zwx}. Instead
of this, we can regularize `formally' the $p$-adic Koba-Nielsen
amplitudes by expressing them as a sum of local zeta functions,
without using the fact that all these functions are holomorphic in a
common domain, fact that was established in
\cite{Bocardo-Gaspar:2016zwx}.

The article is organized in the following form. In Section 2, we
provide a brief review of the limit $p\to 1$ in the effective action
following the results from  \cite{Gerasimov:2000zp}. In particular,
we emphasize that in this limit, the theory with a logarithmic
potential, given in \cite{Gerasimov:2000zp}, gives rise to Feynman
rules that by definition generate Feynman tree-level amplitudes of
the $p$-adic open string in the limit $p \to 1$. Section 3 will be
devoted to present the extension of our results
\cite{Bocardo-Gaspar:2016zwx} to unramified finite field extensions
of the $p$-adic field. Section 4 gives the description of the $p \to
1$ limit of the $p$-adic string amplitudes. For this we use the
formulation of topological zeta functions
\cite{denefandloeser,trossmann}. We also present the computation of
$N=4$ and $N=5$ points Denef-Loeser amplitudes. In Section 5, we
give some final comments. In appendices A and B at the end of the
article, we review some mathematical results employed along Sections
3 and 4.

\section{The limit $p \to 1$ in the effective action}
\label{sec:REV}

In this section we will briefly overview some of the results from
\cite{Gerasimov:2000zp}.  As we mentioned before in
\cite{Brekke:1988dg}, it was argued than the effective action on the
$D$-dimensional target spacetime $M$ and from which one can obtain
the $p$-adic scattering amplitudes at tree-level is given by
\begin{equation}
S(\phi) = {1 \over g^2} {p^2\over p-1} \int d^Dx \bigg(-{1 \over 2}
\phi p^{-{1 \over 2} \Delta} \phi + {1 \over p+1} \phi^{p+1} \bigg),
\label{goodaction}
\end{equation}
where $g$ is the coupling constant, $\Delta$ is the Laplacian on the
underlying spacetime $M$ and $D$ is the dimension of $M$, which is,
in principle, arbitrary. The equation of motion is
\begin{equation}
p^{-{1 \over 2} \Delta} \phi = \phi^p. \label{eomgeneral}
\end{equation}
This equation has different solitonic solutions depending of the
value of $p$ \cite{Sen:1999xm,Moeller:2002vx,Fuchs:2008cc}.

Remember that $p$ is a prime number, which is a parameter in the
equation of motion (\ref{eomgeneral}), since this equation is
formulated in the target space $\mathbb{R}^D$, we can extend $p$ to
be a real parameter.

By considering formally that $p$ is a real variable and comparing
the Taylor expansion of $\exp(-{1\over 2} \Delta \log p)$ and
$\exp(p\log \phi)$ at $(p-1)$, we get that the equation of motion
(\ref{eomgeneral}) becomes
\begin{equation}
\Delta \phi = - 2 \phi \log \phi, \label{eomlineal}
\end{equation}
which can be interpreted as a `linearization' of (\ref{eomgeneral})
in the variable $p$. This is a linear theory with potential
\begin{equation}
V(\phi) =  \phi^2 \log {\phi^2 \over e}.
\end{equation}
Thus `the $p\to 1$ limit of effective action' yields
\begin{equation}
S(\phi) = \int d^Dx\bigg((\partial \phi)^2 - V(\phi) \bigg),
\label{linearaction}
\end{equation}
where $(\partial \phi)^2 = \eta^{ij} \partial_i \phi \cdot
\partial_j \phi$ and $\eta^{ij}$ is the inverse of Minkowski metric
\begin{equation}
\eta_{ij} = {\rm diag}(-1,1,\dots ,1),
\end{equation}
in the sense that action (\ref{linearaction}) leads to equations of
motion (\ref{eomlineal}). Notice that the factor ${p^2 \over g^2
(p-1)}$ in action (\ref{goodaction}) does not play any role in the
linearization of the equation of motion (\ref{eomgeneral}) around
$p=1$. The computation of the correlation functions of the
interacting theory can be done leaving out the mentioned factor. At
the end of the computation the coupling constant $g$ can be
introduced again without any problem.

Then Feynman rules which can be derived from the above Lagrangian
are simple to obtain (see for instance \cite{Peskin:1995ev}). The
free theory with a source term is given by
\begin{equation}
S_0(\phi)= \int d^Dx\big[(\partial \phi)^2 + \phi^2(x) + J(x)\phi(x)
\big].
\end{equation}
The equation of motion is given by
\begin{equation}
(\Delta -1)\phi(x) = {1 \over 2}J(x).
\end{equation}
We use the following notation and conventions:
$$
\phi(x) = -\int d^Dy G(x,y) {J(y) \over 2}, \ \ \ \ \ G(x,y)=\int
{d^Dk \over (2 \pi)^D} e^{i {\boldsymbol{k} \cdot (x-y)}}
G(\boldsymbol{k})
$$
and
$$
\delta^D(x-y)=\int {d^Dk \over (2 \pi)^D} e^{i \boldsymbol{k} \cdot
(x-y)},
$$
where $G(x,y)$ is the Green function of operator $\Delta -1$ and
$G(\boldsymbol{k})$ is its Fourier transform. After a standard
analysis in quantum field theory one finds that the propagator
$x_{ij}$, represent the Green function $G(\boldsymbol{k})$, and can
be expressed as
\begin{equation}
x_{ij}= {1 \over {\boldsymbol{k}_{i} \cdot \boldsymbol{k}_{j}} +1},
\label{Fpropagator}
\end{equation}
where we are using the notation for the propagator from
\cite{Brekke:1988dg}. Here $\boldsymbol{k}_{i}$ with $i=1,\dots,N$
are the external momenta of the scattered particles. The products
$\boldsymbol{k}_{i} \cdot \boldsymbol{k}_{j}$ for all the possible
values of pairs $i,j$ represent the different tachyons propagating
in channels $s$, $t$ and $u$. Moreover, the interactions are
represented by vertices with four external lines attached to each
vertex.

In \cite{Gerasimov:2000zp} it was argued that action
(\ref{linearaction}) equivalently describes the tree-level of the
tachyon field (without quantum corrections) and neglecting all other
fields of the BSFT action given in
\cite{Witten:1992qy,Witten:1992cr}. The relation is performed
through a simple field redefinition $T= - \log \phi^2$, where $T$ is
the tachyon field.

\subsection{Amplitudes from the Gerasimov-Shatashvili Lagrangian}

In this section we show how to extract the four and five-point
amplitudes of the Gerasimov-Shatashvili Lagrangian
(\ref{linearaction}) found in \cite{Gerasimov:2000zp}. In order to
do that, we first require to study the interacting theory. The
generating functional of the correlation function for the free
theory is given by
\begin{equation}
{\cal Z}_0[J] = {\cal N} [\det(\Delta -1)]^{-1/2}\exp
\bigg\{-{i\over 4 \hbar} \int d^Dx \int d^Dx' J(x) G_F(x-x') J(x')
\bigg\},
\end{equation}
where $G_F(x-x')$ is the Green-Feynman function of time-ordered
product of two fields of the theory, ${\cal N}$ is a normalization
constant, $[\det(\Delta -1)]^{-1/2}$ is a suitable regularization of
the divergent determinant bosonic operator, see e.g.
\cite{Peskin:1995ev}.

Action (\ref{linearaction}) can be conveniently rewritten as
\begin{equation}
S(\phi) = \int d^D x \big[(\partial \phi)^2 + m^2 - U(\phi) \big],
\label{actionU}
\end{equation} where $U(\phi) = 2\phi^2 \log \phi$.
We expand $U(\phi)$ in Taylor series around the origin as follows:
\begin{equation}
U(\phi) = A \phi^2 + B \phi^3 + C\phi^4 + D\phi^5 + \cdots ,
\end{equation}
where $A,B,C$ and $D$ are certain real constants.

In the standard formalism of QFT \cite{Peskin:1995ev}, the $N$-point
correlation functions are proportional to
\begin{equation}
 \langle T(\widehat{\phi}(x_1) \widehat{\phi}(x_2) \cdots
\widehat{\phi}(x_N)) \rangle = {(-i \hbar)^N \over {\cal Z}[J]}
{\delta^n {\cal Z}[J] \over \delta J(x_1) \delta J(x_2) \cdots
\delta J(x_N)} \bigg|_{J=0},
\end{equation}
where the $\widehat{\phi}$'s are $N$ local operators (observables)
in $N$ different points $x_1,x_2, \dots , x_N$ of the Minkowski
spacetime, ${\cal Z}[J]$ is the generating functional constructed
using inte\-racting Lagrangian (\ref{actionU}). The functional can
be computed as
$$
{\cal Z}[J] = \exp \bigg\{-{iB \over \hbar} \int d^Dx \bigg(-i \hbar
{\delta \over \delta J(x)} \bigg)^3
$$
\begin{equation}
 -{iC \over \hbar} \int d^Dx
\bigg(-i \hbar {\delta \over \delta J(x)} \bigg)^4  -{iD \over
\hbar} \int d^Dx \bigg(-i \hbar {\delta \over \delta J(x)} \bigg)^5
+ \cdots \bigg\} {\cal Z}_0[J].
\end{equation}
We assert that connected tree-level scattering amplitudes of this
theory match exactly with the corresponding amplitudes of the
effective action (\ref{goodaction}) in the limit when $p$ tends to
one.

\subsection{Four-point amplitudes}

The 4-point amplitudes can be computed as follows: the 4-point
vertex can be obtained purely from the quartic interaction at the
first order in perturbation theory. The generating functional, with
the vertex labeled by $x$ and 4 external legs attached to it, is
given by
\begin{equation}
{\cal Z}[J] = \cdots -i C \hbar^3 \int d^Dx\bigg({\delta \over
\delta J(x)}\bigg)^4 {\cal Z}_0[J] + \cdots .
\end{equation}
The corresponding 4-point amplitude is proportional to
$$
{\delta^4 {\cal Z}[J] \over \delta J(x_1) \delta J(x_2)  \delta
J(x_3)  \delta J(x_4)} \bigg|_{J=0}
$$
$$
 = - 4!iC\hbar^3 \int d^D x
\bigg[-{i \over 2 \hbar} G_F(x-x_1) \bigg] \bigg[-{i \over 2 \hbar}
G_F(x-x_2) \bigg] \bigg[-{i \over 2 \hbar} G_F(x-x_3) \bigg]
\bigg[-{i \over 2 \hbar} G_F(x-x_4) \bigg]
$$
\begin{equation}
= - {3iC \over 2 \hbar}  \int d^D x  \ G_F(x-x_1) \ G_F(x-x_2) \
G_F(x-x_3) \ G_F(x-x_4), \label{vertex}
\end{equation}
where $G_F(x-y)$ is the Green-Feynman propagator. In the Fourier
space the above amplitude corresponds to the Feynman diagram with
only one vertex and four external legs. In analogy to the notation
from \cite{Brekke:1988dg}, we will represent it by the letter
$\overline{K}_4$.

The interaction term $B\phi^3$ in the Lagrangian has also a
non-vanishing contribution to the 4-points tree amplitudes at the
second order in perturbation theory. They are described by Feynman
diagrams with two vertices located at points $x$ and $y$ connected
by a propagator $G_F(x-y)$ and with two external legs attached to
each vertex. In this case the amplitude is computed from the
relevant part of the generating functional
\begin{equation}
{\cal Z}[J] = \cdots + {B^2 \hbar^4 \over 2} \int d^Dx \int d^Dy \
\bigg({\delta \over \delta J(x)}\bigg)^3  \bigg({\delta \over \delta
J(y)}\bigg)^3 {\cal Z}_0[J] + \cdots . \label{cubiccubic}
\end{equation}
The connected 4-point amplitudes at the second order of the cubic
interaction $C \phi^3$ yields
$$
{\delta^4 {\cal Z}[J] \over \delta J(x_1) \delta J(x_2)  \delta
J(x_3)  \delta J(x_4)} \bigg|_{J=0} =  18 B^2 \hbar^4\int d^Dx \int
d^Dy \ \bigg[-{i \over 2 \hbar} G_F(x-y) \bigg]
$$
$$
\times \bigg\{ \bigg[-{i \over 2 \hbar} G_F(x-x_4) \bigg] \bigg[-{i
\over 2 \hbar} G_F(x-x_3) \bigg] \bigg[-{i \over 2 \hbar} G_F(y-x_2)
\bigg]  \bigg[-{i \over 2 \hbar} G_F(y-x_1) \bigg]
$$
$$
+ \bigg[-{i \over 2 \hbar} G_F(x-x_4) \bigg] \bigg[-{i \over 2
\hbar} G_F(y-x_3) \bigg] \bigg[-{i \over 2 \hbar} G_F(x-x_2) \bigg]
\bigg[-{i \over 2 \hbar} G_F(y-x_1) \bigg]
$$
$$
+ \bigg[-{i \over 2 \hbar} G_F(x-x_4) \bigg] \bigg[-{i \over 2
\hbar} G_F(y-x_3) \bigg] \bigg[-{i \over 2 \hbar} G_F(y-x_2) \bigg]
\bigg[-{i \over 2 \hbar} G_F(x-x_1) \bigg]
$$
$$
+ \bigg[-{i \over 2 \hbar} G_F(y-x_4) \bigg] \bigg[-{i \over 2
\hbar} G_F(y-x_3) \bigg] \bigg[-{i \over 2 \hbar} G_F(x-x_2) \bigg]
\bigg[-{i \over 2 \hbar} G_F(x-x_1) \bigg]
$$
$$
+ \bigg[-{i \over 2 \hbar} G_F(y-x_4) \bigg] \bigg[-{i \over 2
\hbar} G_F(x-x_3) \bigg] \bigg[-{i \over 2 \hbar} G_F(y-x_2) \bigg]
 \bigg[-{i \over 2 \hbar} G_F(x-x_1) \bigg]
$$
\begin{equation}
+ \bigg[-{i \over 2 \hbar} G_F(y-x_4) \bigg] \bigg[-{i \over 2
\hbar} G_F(x-x_3) \bigg] \bigg[-{i \over 2 \hbar} G_F(x-x_2) \bigg]
\bigg[-{i \over 2 \hbar} G_F(y-x_1) \bigg] \bigg\}. \label{segunda}
\end{equation}
This amplitude corresponds to the scattering of particles
propagating in the sum of the $s$, $t$ and $u$ channels. They
together with the 4-point vertex (\ref{vertex}) constitute the
tree-level amplitudes arising in the 4-point $p$-adic amplitudes in
the limit when $p \to 1$. Thus in the Fourier space the total
amplitude for 4-point amplitudes consists of the sum of the
amplitude given by Eq. (\ref{vertex})  plus the contribution
(\ref{segunda}) that we schematically write (in notation from
\cite{Brekke:1988dg}) as
\begin{equation}
A_4 = \overline{K}_4 + \sum_{i<j} x_{ij},
\end{equation}
where $x_{ij}$ is given by (\ref{Fpropagator}).

\subsection{Five-point amplitudes}

For the 5-point amplitudes there is a contribution coming from the
quintic interaction term $D\phi^5$ in the Lagrangian. Thus we have
at the first order in the perturbative expansion that the relevant
contribution of the generating functional is given by
\begin{equation}
{\cal Z}[J] = \cdots -{D  \hbar^4} \int d^Dx\bigg( {\delta \over
\delta J(x)}\bigg)^5 {\cal Z}_0[J] + \cdots .
\end{equation}
The vertex function for the 5-point amplitude reads
$$
{\delta^5 {\cal Z}[J] \over \delta J(x_1) \delta J(x_2)  \delta
J(x_3)  \delta J(x_4) \delta J(x_5)} \bigg|_{J=0} =  - 5! D \hbar^4
\int d^D x \bigg[-{i \over 2 \hbar} G_F(x-x_1) \bigg]
$$
\begin{equation}
\times \bigg[-{i \over 2 \hbar} G_F(x-x_2) \bigg] \bigg[-{i \over 2
\hbar} G_F(x-x_3) \bigg] \bigg[-{i \over 2 \hbar} G_F(x-x_4) \bigg]
\bigg[-{i \over 2 \hbar} G_F(x-x_5) \bigg]. \label{unouno}
\end{equation}
Similarly to the case of 4-point amplitudes, the above amplitude is
represented by a diagram with only one vertex and five external legs
and the amplitude denoted by $\overline{K}_5$ in the Fourier space.

Now we study the possible terms to the 5-point tree-amplitude coming
from the interaction term $B \phi^3 \times C \phi^4$. This term
consist of $p$-adic amplitudes in the fourier space constructed from
amplitudes with 2-vertices, 5 external legs and one internal leg as
described in Sec. 3 from \cite{Brekke:1988dg}.

The relevant part of the generating functional is given by
\begin{equation}
{\cal Z}[J] = \cdots - iBC \hbar^5 \int d^Dx \int d^Dy \
\bigg({\delta \over \delta J(x)}\bigg)^3  \bigg({\delta \over \delta
J(y)}\bigg)^4 {\cal Z}_0[J] + \cdots . \label{cubicquartic}
\end{equation}
The computation of a 5-point amplitude from this generating
functional is given by
$$
{\delta^5 {\cal Z}[J] \over \delta J(x_1) \delta J(x_2)  \delta
J(x_3)  \delta J(x_4) \delta J(x_5)} \bigg|_{J=0}
$$
$$
= {-iBC (12)^2 \hbar^5} \int d^D x  \int d^D y \ \bigg[-{i \over 2
\hbar} G_F(x-y) \bigg] \bigg\{ \bigg[-{i \over 2 \hbar} G_F(x-x_5)
\bigg]\bigg[-{i \over 2 \hbar} G_F(x-x_4) \bigg]
$$
$$
\times \bigg[-{i \over 2 \hbar} G_F(y-x_3) \bigg] \bigg[-{i \over 2
\hbar} G_F(y-x_2) \bigg] \bigg[-{i \over 2 \hbar} G_F(y-x_1) \bigg]
+ \cdots
$$
$$
+ \bigg[-{i \over 2 \hbar} G_F(y-x_5) \bigg] \bigg[-{i \over 2
\hbar} G_F(x-x_4) \bigg]   \bigg[-{i \over 2 \hbar} G_F(x-x_3)
\bigg]
$$
\begin{equation}
\times\bigg[-{i \over 2 \hbar} G_F(y-x_2) \bigg] \bigg[-{i \over 2
\hbar} G_F(y-x_1) \bigg] + \cdots \bigg\}. \label{fivepointstwov}
\end{equation}

There will be in total ten terms in equation (\ref{fivepointstwov})
including the possible permutations of labels $(x_1,\dots ,x_5)$ and
of the two vertices at $x$ and $y$.

Finally the lacking contribution to the 5-point amplitudes comes
from the third order of the cubic interaction term in the
Lagrangian.  We have three vertices labeled by $x$, $y$ and $z$. Two
of these vertices are connected to two external legs and to one
internal line.  The other is attached to two internal lines and one
external. Thus the generating function in this case is given by
\begin{equation}
{\cal Z}[J] = \cdots + {B^3 \hbar^6 \over 3!} \int d^Dx \int d^Dy
\int d^Dz \ \bigg({\delta \over \delta J(x)}\bigg)^3 \cdot
\bigg({\delta \over \delta J(y)}\bigg)^3 \cdot \bigg({\delta \over
\delta J(z)}\bigg)^3{\cal Z}_0[J] + \cdots . \label{cubiccubiccubic}
\end{equation}
The contribution of these terms to the 5-point function results
$$
{\delta^5 {\cal Z}[J] \over \delta J(x_1) \delta J(x_2)  \delta
J(x_3)  \delta J(x_4) \delta J(x_5)} \bigg|_{J=0}
 =  {B^3 \hbar^6 a \over 3!}  \int d^D x  \int d^D y \int d^D z \
\bigg[-{i \over 2 \hbar} G_F(x-y) \bigg]
$$
$$
\times  \bigg[-{i \over 2 \hbar} G_F(y-z) \bigg]\bigg\{ \bigg[-{i
\over 2 \hbar} G_F(y-x_5) \bigg] \bigg[-{i \over 2 \hbar} G_F(z-x_4)
\bigg]\bigg[-{i \over 2 \hbar} G_F(z-x_3) \bigg]
$$
\begin{equation}
\times \bigg[-{i \over 2 \hbar} G_F(x-x_2) \bigg] \bigg[-{i \over 2
\hbar} G_F(x-x_1) \bigg] + \cdots \bigg\},
\label{tres}
\end{equation}
where $a$ is a suitable constant.

Thus we summarize the contributions to the 5-point amplitudes $A_5$
obtained from the Gerasimov and Shatashvili Lagrangian. This is
written schematically by
$$
A_5 =\overline{K}_5 + \sum_{i<j} \overline{K}_4 x_{ij} +
\sum_{i<j}\sum_{k<l} x_{ij} \cdot x_{kl},
$$
where the three terms in the sum correspond to the amplitudes given
by Eqs. (\ref{unouno}), (\ref{fivepointstwov}) and (\ref{tres})
respectively, and where $x_{ij}$ is given by (\ref{Fpropagator}).

Through out this procedure we can compute a $N$-point tree-level
$p$-adic string amplitude in the limit $p\to 1$ for any number of
external legs $N$. As we have argued in this section, this amplitude
can be obtained from the Gerasimov-Shatashvili action with a
logarithmic potential (\ref{linearaction}).

Notice that the calculations involving the limit $p \to 1$ in the
case of effective action are performed in $\mathbb{R}^{D}$,
meanwhile the calculations involving the limit $p \to 1$ in the case
of $p$-adic string amplitudes are performed in  $\mathbb{Q}_p^{D}$,
and in the $p$-adic topology the limit $p \to 1$ does not make
sense. In section 4 we will give a rigorous procedure to get the
limit $p \to 1$ in the amplitudes and we will reproduce the correct
Feynman rules discussed in the present section. As a byproduct one
can see that these amplitudes can be computed in a more economic and
efficient way by using this rigorous procedure.


\section{Koba-Nielsen string amplitudes on finite extensions of non-Archimedean local fields}

\subsection{$p$-adic string amplitudes}

In \cite{Brekke:1988dg} Brekke et al discussed the amplitudes for
the $N$-point tree-level $p$-adic bosonic string amplitudes. They
also computed the four and five-points amplitudes explicitly and it
was investigated how these amplitudes can be obtained from an
effective Lagrangian. The open string $N-$point tree amplitudes over
the $p$-adic field $\mathbb{Q}_{p}$ are defined as
\begin{equation}
\boldsymbol{A}^{(N)}\left(  \underline{\boldsymbol{k}}\right) =
\int\limits_{\mathbb{Q}_{p}^{N-3}} \prod\limits_{i=2}^{N-2}
\left\vert x_{i}\right\vert _{p}^{\boldsymbol{k}_{1} \cdot
\boldsymbol{k}_{i} }\left\vert
1-x_{i}\right\vert_{p}^{\boldsymbol{k}_{N-1} \cdot
\boldsymbol{k}_{i}} \ \prod\limits_{2\leq i<j\leq N-2} \left\vert
x_{i}-x_{j}\right\vert _{p}^{\boldsymbol{k}_{i} \cdot
\boldsymbol{k}_{j}} \ \prod\limits_{i=2}^{N-2} dx_{i},
\label{Amplitudeone}
\end{equation}
where $|\cdot |_p$ is the $p$-adic norm (see appendix A),
$\prod\nolimits_{i=2}^{N-2} dx_{i}$ is the normalized Haar measure
of $\mathbb{Q}_{p}^{N-3}$, $\underline {\boldsymbol{k}}=\left(
\boldsymbol{k}_{1},\ldots,\boldsymbol{k}_{N}\right)$ and
$\boldsymbol{k}_{i}=\left(  k_{0,i},\ldots,k_{25,i}\right),$
$i=1,\ldots,N$, $N\geq4$, are the momentum components of the $i$-th
tachyon (with Minkowski inner product $\boldsymbol{k}_{i} \cdot \boldsymbol{k}_{j}=-k_{0,i}%
k_{0,j}+k_{1,i}k_{1,j}+\cdots+k_{25,i}k_{25,j}$) obeying
\begin{equation}
\sum_{i=1}^{N}\boldsymbol{k}_{i}=\boldsymbol{0}, \ \ \ \ \ \ \ \
\boldsymbol{k}_{i} \cdot \boldsymbol{k}_{i}=2,
\end{equation}
for $i=1,\ldots,N$. A central problem in string theory is to know
whether integrals of type (\ref{Amplitudeone}) converge for some
complex values $\boldsymbol{k}_{i} \cdot \boldsymbol{k}_{j}$. Our
results in \cite{Bocardo-Gaspar:2016zwx} allow us to solve this
problem.

\subsection{Non-Archimedean String Zeta Functions}

In this subsection we extend some results of our previous work
\cite{Bocardo-Gaspar:2016zwx} from $\mathbb{Q}_{p}$ to
$\mathbb{K}_e$, the unique unramified extension of $\mathbb{Q}_{p}$
of degree $e$. In this article we use most of the notation and
conventions introduced in \cite{Bocardo-Gaspar:2016zwx}.

For a discussion about non-Archimedean local fields, the reader may
consult appendix A or references \cite{We,Taibleson,Alberio et
al,V-V-Z}.

We consider $\mathbb{K}$ a non-Archimedean local field of
characteristic zero. Denote by $R_{\mathbb{K}}$ the {\it ring of
integers} of $\mathbb{K}$, this ring contains a unique maximal ideal
$P_{\mathbb{K}}$, which is principal. We fix a generator $\pi$ (also
called a {\it uniformizing parameter} of $\mathbb{K}$), so
$P_{\mathbb{K}} = \pi R_{\mathbb{K}}$.

Any finite extension $\mathbb{K}$ of $\mathbb{Q}_{p}$ is a
non-Archimedean local field. Then
\begin{equation}
pR_{\mathbb{K}}=\pi^{m} R_{\mathbb{K}}, \ \ \ \ \ m\in \mathbb{N}.
\end{equation}
If $m=1$ we say that $\mathbb{K}$ is a {\it unramified} extension of
$\mathbb{Q}_{p},$ otherwise we say that $\mathbb{K}$ is a {\it
ramified} extension. It is well known that for every positive
integer $e$ there exist a unique unramified extension $\mathbb{K}_e$
of $\mathbb{Q}_{p}$ of degree $e$, which means that $\mathbb{K}_e$
is a $\mathbb{Q}_{p}$-vector space of dimension $e$. From now on,
$\pi$ stands for a local uniformizing parameter of $\mathbb{K}_e$,
thus $pR_{\mathbb{K}_{e}}=\pi R_{\mathbb{K}_{e}}$,
$R_{\mathbb{K}_{e}}/P_{\mathbb{K}_{e}} \cong \mathbb{F}_{p^{e}}$ and
$|\pi|_{\mathbb{K}_e} = p^{-e}$. Thus $\pi$ in $\mathbb{K}_{e}$
plays the role of $p$ in $\mathbb{Q}_p$.

We now describe the generalization of $p$-adic Koba-Nielsen
amplitudes. These amplitudes are generalized as follows:
\begin{equation}
\boldsymbol{A}^{(N)}\left(  \underline{\boldsymbol{k}},
\mathbb{K}_e\right) = \int\limits_{\mathbb{K}_{e}^{N-3}}
\prod\limits_{i=2}^{N-2} \left\vert x_{i}\right\vert
_{\mathbb{K}_e}^{\boldsymbol{k}_{1} \cdot  \boldsymbol{k}_{i}
}\left\vert 1-x_{i}\right\vert _{\mathbb{K}_e}^{\boldsymbol{k}_{N-1}
\cdot \boldsymbol{k}_{i}} \ \prod\limits_{2\leq i<j\leq N-2}
\left\vert x_{i}-x_{j}\right\vert
_{\mathbb{K}_e}^{\boldsymbol{k}_{i} \cdot \boldsymbol{k}_{j}} \
\prod\limits_{i=2}^{N-2} dx_{i} \label{Amplitude}
\end{equation}
where $\prod\nolimits_{i=2}^{N-2} dx_{i}$ is the normalized Haar
measure of $\mathbb{K}_{e}^{N-3}$.

Following \cite{Bocardo-Gaspar:2016zwx}, in order to study the
amplitude
$\boldsymbol{A}^{(N)}\left(\underline{\boldsymbol{k}};\mathbb{K}_{e}\right)$,
we introduce the {\it open string $N$-point zeta function}, which is
defined by
\begin{equation}
\boldsymbol{Z}^{(N)}\left(
\underline{\boldsymbol{s}},\mathbb{K}_{e}\right)  :=
\int_{{\mathbb{K}_{e}}^{N-3}{\smallsetminus\Lambda}}
F\left(\underline{\boldsymbol{s}},\boldsymbol{x};N,\mathbb{K}_{e}\right)
{\displaystyle\prod\limits_{i=2}^{N-2}} dx_{i}, \label{Zeta_1}
\end{equation}
where
\begin{equation}
F\left(
\underline{\boldsymbol{s}},\boldsymbol{x};N,\mathbb{K}_{e}\right)  =
{\displaystyle\prod\limits_{i=2}^{N-2}} \left\vert x_{i}\right\vert
_{\mathbb{K}_{e}}^{s_{1i}}\left\vert 1-x_{i}\right\vert
_{\mathbb{K}_e}^{s_{(N-1)i}}\text{ }  \
{\displaystyle\prod\limits_{2\leq i<j\leq N-2}} \left\vert
x_{i}-x_{j}\right\vert _{\mathbb{K}_{e}}^{s_{ij}}.
\end{equation}
Here we assume that $\underline{\boldsymbol{s}}=\left(s_{ij}\right)
\in\mathbb{C}^{D}$, with $D=\frac{\left( N-3\right) \left(
N-4\right)}{2}+2\left( N-3\right)$. Moreover $s_{ij}=s_{ji}$ for all
$i,j$ and $\boldsymbol{x}=\left( x_{2},\ldots,x_{N-2}\right) \in
\mathbb{K}_{e}^{N-3}$ and $\Lambda$ is defined by
\begin{equation}
\Lambda:=\left\{  \left(  x_{2},\ldots,x_{N-2}\right)  \in
\mathbb{K}_{e}^{N-3}; {\displaystyle\prod\limits_{i=2}^{N-2}}
x_{i}\left(  1-x_{i}\right)  \text{ } \
{\displaystyle\prod\limits_{2\leq i<j\leq N-2}} \left(
x_{i}-x_{j}\right)  =0\right\}
\end{equation}
and ${\textstyle\prod\nolimits_{i=2}^{N-2}} dx_{i}$ is the Haar
measure of $\mathbb{K}_{e}^{N-3}$ normalized so that the measure of
$R_{\mathbb{K}_{e}}^{N-3}$ is $1$. The name for
$\boldsymbol{Z}^{(N)}\left(
\underline{\boldsymbol{s}},\mathbb{K}_{e}\right)$ comes from the
fact that it is a finite sum of multivariate local zeta functions,
as we explain below, see also Appendix B. In the definition of Eq.
(\ref{Zeta_1}) we remove the set $\Lambda$ from the domain of
integration in order to use the formula $a^s= e^{s \ln a}$ for $a>0$
and $s \in \mathbb{C}$.

For a subset $I$ of $T=\{2,\dots,N-2\}$, we define the zeta function
\begin{equation}
\boldsymbol{Z}^{(N)}\left(
\underline{\boldsymbol{s}};I,\mathbb{K}_{e}\right)=
{\displaystyle\int\limits_{Sect(I)}} F\left(
\underline{\boldsymbol{s}},\boldsymbol{x};N,\mathbb{K}_{e}\right)
{\displaystyle\prod\limits_{i=2}^{N-2}} dx_{i},
\end{equation}
attached to the sector
\begin{equation}
Sect(I)=\left\{  \left(  x_{2},\ldots,x_{N-2}\right)  \in \mathbb{K}_{e}^{N-3}%
;\left\vert x_{i}\right\vert _{\mathbb{K}_{e}}\leq 1 \Leftrightarrow
i\in I\right\}.
\end{equation}

Then
$\boldsymbol{Z}^{(N)}\left(\underline{\boldsymbol{s}},\mathbb{K}_{e}\right)$
is a sum over all the possible inequivalent sectors $Sect(I)$:
\begin{equation}
\boldsymbol{Z}^{(N)}\left(
\underline{\boldsymbol{s}},\mathbb{K}_{e}\right) =\sum_{I\subseteq
T} \boldsymbol{Z}^{(N)}\left(
\underline{\boldsymbol{s}};I,\mathbb{K}_{e}\right).
\end{equation}
As in \cite{Bocardo-Gaspar:2016zwx}, we can show that
\begin{equation}
\boldsymbol{Z}^{(N)}\left(
\underline{\boldsymbol{s}},\mathbb{K}_{e}\right) =\sum_{I\subseteq
T}p^{eM(\underline{\boldsymbol{s}})} \boldsymbol{Z}^{(N)}\left(
\underline{\boldsymbol{s}};I,0,\mathbb{K}_{e}\right) \
\boldsymbol{Z}^{(N)}\left( \underline{\boldsymbol{s}};
T\smallsetminus I,1,\mathbb{K}_{e}\right),
\label{Formula_zeta_amplitude}
\end{equation}
where
\begin{equation}
M(\underline{\boldsymbol{s}}):=\left\vert T\smallsetminus I
\right\vert + \sum_{i\in T\smallsetminus I}(s_{1i}+s_{\left(
N-1\right) i} )+\sum_{\substack{2\leq i<j\leq N-2\\i\in
T\smallsetminus I, \ j\in T} }s_{ij}+\sum_{\substack{2\leq i<j\leq
N-2\\i\in I,j\in T\smallsetminus I}}s_{ij}. \label{exponente}
\end{equation}

The functions $\boldsymbol{Z}^{(N)}\left(
\underline{\boldsymbol{s}};I,0,\mathbb{K}_{e}\right)$ and
$\boldsymbol{Z}^{(N)}\left(
\underline{\boldsymbol{s}};T\smallsetminus
I,1,\mathbb{K}_{e}\right)$ are given by
\begin{equation}
\boldsymbol{Z}^{(N)}\left(
\underline{\boldsymbol{s}};I,0,\mathbb{K}_{e}\right) =
{\displaystyle\int\limits_{R_{\mathbb{K}_{e}}^{\left\vert
I\right\vert }}} F_0\left(
\boldsymbol{s},\boldsymbol{x};N,\mathbb{K}_{e}\right)
 \  {\displaystyle\prod\limits_{i\in I}} dx_{i},
\end{equation}
where
\begin{equation}
F_0\left(
\underline{\boldsymbol{s}},\boldsymbol{x};N,\mathbb{K}_{e}\right) :=
{\displaystyle\prod\limits_{i\in I}} \left\vert x_{i}\right\vert
_{\mathbb{K}_{e}}^{s_{1i}}\left\vert 1-x_{i}\right\vert
_{\mathbb{K}_{e}}^{s_{(N-1)i}}\text{ } \
{\displaystyle\prod\limits_{\substack{2\leq i<j\leq N-2\\i,j\in I}}}
\left\vert x_{i}-x_{j}\right\vert _{\mathbb{K}_{e}}^{s_{i}{}_{j}}
\end{equation}
and
\begin{equation}
\boldsymbol{Z}^{(N)}\left(
\underline{\boldsymbol{s}};T\smallsetminus
I,1,\mathbb{K}_{e}\right)=
{\displaystyle\int\limits_{R_{\mathbb{K}_{e}}^{\left\vert
T\smallsetminus I\right\vert }}} F_1\left(
\underline{\boldsymbol{s}},\boldsymbol{x};N,\mathbb{K}_{e}\right) \
{\displaystyle\prod\limits_{i\in T\smallsetminus I}} dx_{i},
\end{equation}
where
\begin{equation}
F_1\left(
\underline{\boldsymbol{s}},\boldsymbol{x};N,\mathbb{K}_{e}\right) :=
\frac{ {\displaystyle\prod\limits_{\substack{2\leq i<j\leq
N-2\\i,j\in T\smallsetminus I}}} \left\vert x_{i}-x_{j}\right\vert
_{\mathbb{K}_{e}}^{s_{ij}}}{ {\displaystyle\prod\limits_{i\in
T\smallsetminus I}} \left\vert x_{i}\right\vert
_{\mathbb{K}_{e}}^{2+s_{1i}+s_{\left( N-1\right)i} +\sum_{2\leq
j\leq N-2,j\neq i}s_{ij}}}.
\end{equation}
By convention $\boldsymbol{Z}^{(N)}\left(
\underline{\boldsymbol{s}};\varnothing,0,\mathbb{K}_{e}\right) =1$,
$\boldsymbol{Z}^{(N)}\left( \underline
{\boldsymbol{s}};\varnothing,1,\mathbb{K}_{e}\right) =1$. Regarding
the notation, for $J\subseteq T$, $J\not=\varnothing$, we denote by
$R^{|J|}_{\mathbb{K}_e}$ the set $\{(x_i)_{i\in J}; x_i \in
R_{\mathbb{K}_e}\}$, if $J = \varnothing$, then
$R^{|J|}_{\mathbb{K}_e}=\varnothing$. We denote by $T\smallsetminus
I= \{ j\in T; j \notin I \}$.

The functions $\boldsymbol{Z}^{(N)}\left(
\underline{\boldsymbol{s}};I,0,\mathbb{K}_{e}\right)$ and
$\boldsymbol{Z}^{(N)}\left(
\underline{\boldsymbol{s}};T\smallsetminus
I,1,\mathbb{K}_{e}\right)$ are multivariate local zeta functions,
see Apendix B. The local zeta functions are related with deep
arithmetical and geometrical matters, and they have been studied
extensively since the 50´s, see \cite{Meuser,Denef} and references
therein.

In \cite{Bocardo-Gaspar:2016zwx} we showed that
$\boldsymbol{Z}^{(N)}\left( \underline{\boldsymbol{s}},
\mathbb{Q}_p\right)$ has an analytic continuation to the whole
$\mathbb{C}^{D}$ as a rational function in the variables
$p^{-s_{ij}}$, see Propositions 1, 2 and Theorem 1 in
\cite{Bocardo-Gaspar:2016zwx}. These results are valid for finite
extensions of $\mathbb{Q}_p$. More precisely, all the zeta functions
appearing in the right-hand side of formula
(\ref{Formula_zeta_amplitude}) admit analytic continuations to the
whole $\mathbb{C}^{D}$ as rational functions in the variables
$p^{-es_{ij}}$. In addition, each of these functions is holomorphic
on a certain domain in $\mathbb{C}^{D}$ and the intersection of all
these domains contains an open and connected subset of
$\mathbb{C}^{D}$. Therefore  $\boldsymbol{Z}^{(N)}\left(
\underline{\boldsymbol{s}}, \mathbb{K}_e\right)$ is a holomorphic
function in a certain domain of $\mathbb{C}^{D}$ admitting a
meromorphic continuation to the whole $\mathbb{C}^{D}$ as a rational
function in the variables  $p^{-es_{ij}}$, see Theorem 1 in
\cite{Bocardo-Gaspar:2016zwx}.

We use $\boldsymbol{Z}^{(N)}(\underline
{\boldsymbol{s}},\mathbb{K}_{e})$ as regularizations of Koba-Nielsen
amplitudes
$\boldsymbol{A}^{(N)}\left(\underline{\boldsymbol{k}},\mathbb{K}_{e}\right)$,
more precisely, we define
\begin{equation}
\boldsymbol{A}^{(N)}\left(
\underline{\boldsymbol{k}},\mathbb{K}_{e}\right)
=\boldsymbol{Z}^{(N)}(\underline{\boldsymbol{s}},\mathbb{K}_{e})\mid_{s_{ij}
=\boldsymbol{k}_{i} \cdot \boldsymbol{k}_{j}}.
\end{equation}
Then $\boldsymbol{A}^{(N)}\left(
\underline{\boldsymbol{k}},\mathbb{K}_{e}\right)$ is a well defined
rational function in the variables  $p^{-e\boldsymbol{k}_{i} \cdot
\boldsymbol{k}_{j}}$, which agree with the integral
(\ref{Amplitude}) when it converges.

\section{The limit $p \to 1$ in $p$-adic string amplitudes}

In the previous sections we have seen that the $p$-adic string
amplitudes are essentially local zeta functions, explicitly
$\boldsymbol{Z}^{(N)}\left(\underline{\boldsymbol{s}};I,0,\mathbb{K}_{e}\right)$
and $\boldsymbol{Z}^{(N)}\left(\underline
{\boldsymbol{s}};T\smallsetminus I,1,\mathbb{K}_{e}\right)$ are both
multivariate local zeta functions of type
$\boldsymbol{Z}\left(\boldsymbol{s},\boldsymbol{f},\mathbb{K}_{e}\right)$
for suitable $\boldsymbol{f}$ (for more details see Appendix B).

\subsection{Topological Zeta functions}

To make mathematical sense of the limit of
$\boldsymbol{Z}^{(N)}\left( \underline{\boldsymbol{s}},
\mathbb{Q}_p\right)$ as $p\to 1$ we use the work of Denef and
Loeser, see \cite{denefandloeser} and \cite{trossmann}. The first
step is to pass from $\mathbb{Q}_p$ to $\mathbb{K}_e$, $e \in
\mathbb{N}$, and consider $\boldsymbol{Z}^{(N)}\left(
\underline{\boldsymbol{s}}, \mathbb{K}_e\right)$ instead of
$\boldsymbol{Z}^{(N)}\left( \underline{\boldsymbol{s}},
\mathbb{Q}_p\right)$, and compute the limit of
$\boldsymbol{Z}^{(N)}\left( \underline{\boldsymbol{s}},
\mathbb{K}_e\right)$ as $e \to 0$ instead of the limit of
$\boldsymbol{Z}^{(N)}\left( \underline{\boldsymbol{s}},
\mathbb{Q}_p\right)$ as $p \to 1$. In order to compute the limit $e
\to 0$ is necessary to have an explicit formula for
$\boldsymbol{Z}^{(N)}\left( \underline{\boldsymbol{s}},
\mathbb{K}_e\right)$ which is equivalent to have explicit formulas
for integrals
$\boldsymbol{Z}^{(N)}\left(\underline{\boldsymbol{s}};I,0,\mathbb{K}_{e}\right)$
and $\boldsymbol{Z}^{(N)}\left(\underline
{\boldsymbol{s}};T\smallsetminus I,1,\mathbb{K}_{e}\right)$, see
(\ref{Formula_zeta_amplitude}). These integrals are special types of
multivariate local zeta functions $\boldsymbol{Z}\left(
\underline{\boldsymbol{s}},f, \mathbb{K}_e\right)$, see Appendix B.
Consequently, we need an explicit formula for the multivariate
Igusa's local zeta function $\boldsymbol{Z}\left(
\underline{\boldsymbol{s}},f, \mathbb{K}_e\right)$, this formula is
a simple variation of the explicit formula established by Denef
\cite{DL3}, which requires Hironaka's desingularization Theorem
\cite{Hi}, see also Appendix B1.

Let $\boldsymbol{f=}\left( f_{1},\ldots ,f_{r}\right) $ with
$f_{i}(\boldsymbol{x})\in \mathbb{Z} \left[ \boldsymbol{x}\right] ,$
$\boldsymbol{x}=\left( x_{1},\ldots ,x_{n}\right),$ be a
non-constant polynomial for $i=1,\ldots ,r$. Let $\left(
Y,\,h\right) $ an embedded resolution of singularities for $D=$ Spec
$\mathbb{Q}\left[\boldsymbol{x}\right]
/\left(\prod\nolimits_{i=1}^{r}f_{i}(\boldsymbol{x})\right)$ over
$\mathbb{Q}$ with $\left\{E_{i}\right\} $ the irreducible components
of $h^{-1}(0)$.

For any scheme $V$ of finite type over a field $L\subset
\mathbb{C}$, we denote by $\chi \left( V\right) $ the Euler
characteristic of the $\mathbb{C} $-analytic space associated with
$V$. Denef and Loeser associated to
$\prod\nolimits_{i=1}^{r}f_{i}(\boldsymbol{x})$ the following
function (the topological zeta function):
\begin{equation}
\boldsymbol{Z}_{top}\left( \boldsymbol{s}\right)
=\sum\limits_{I\subseteq T}\chi \left( \overset{\circ
}{E_{I}}\right) \prod\limits_{i\in
I}\frac{1}{v_{i}+\sum_{j=1}^{r}N_{ij}s_{j}},
\label{Topol_zeta}
\end{equation}
for the notation, see Appendix B.

In arbitrary dimension there is no a canonical way of picking an
embedded resolution of singularities for a divisor. Then, it is
necessary to show that definition (\ref{Topol_zeta}) is independent
of the resolution of singularities chosen, this fact was established
by Denef and Loeser in \cite{denefandloeser}, see also
\cite{trossmann}. By using the explicit formula
(\ref{Explicit_for_1})-(\ref{Explicit_for_2}), Denef and Loeser
showed that
\begin{equation}
\boldsymbol{Z}_{top}\left( \boldsymbol{s}\right) =\lim_{e\to 0}%
\boldsymbol{Z}\left(
\boldsymbol{s},\boldsymbol{f},\mathbb{K}_{e}\right).
\label{limir_p_goes_1}
\end{equation}
The limit $e\to 0$ makes sense because one can $l$-adically
interpolate $\boldsymbol{Z}\left(
\boldsymbol{s},\boldsymbol{f},\mathbb{K}_{e}\right)$ as a function
of $e$. This means that there exist $\kappa \in
\mathbb{N\smallsetminus }\left\{ 0\right\} $ and a meromorphic
function in the variables $\boldsymbol{s}$ and $e$,
$\boldsymbol{Z}_{l}\left( \boldsymbol{s},\boldsymbol{f},e,n\right)$
on $\mathbb{Z}_{l}^{r}\times \left( \kappa \mathbb{Z}_{l}\right) $
such that for any $\boldsymbol{s}\in \mathbb{N}^{r}$ and $e\in
\kappa \mathbb{Z}_{l}$ verifies that
\begin{equation}
\boldsymbol{Z}_{l}\left( \boldsymbol{s},\boldsymbol{f},e,n\right) =%
\boldsymbol{Z}\left(
\boldsymbol{s},\boldsymbol{f},\mathbb{K}_{e}\right) .
\end{equation}
In addition, it is possible to choose $\kappa $ such that
$\boldsymbol{Z}_{l}\left( \boldsymbol{s},\boldsymbol{f},e,n\right)
\left(v_{i}+\sum_{j=1}^{r}N_{ij}s_{j}\right)^{n}$ is a convergent
series on $\mathbb{Z}_{l}^{r}\times \left(
\kappa\mathbb{Z}_{l}\right)$.

In particular
\begin{equation}
\lim_{e\to 0}c_{I}(\mathbb{K}_{e})=\chi _{c}\left( \overset{\circ
}{E_{I}} \otimes \mathbb{F}_{p^{e}}^{a},\mathcal{F}_{\chi
_{triv}}\right) =\chi \left( \overset{\circ }{E_{I}}\right),
\end{equation}
for almost all prime number $p$, where $\chi_{c}$ denotes the Euler
characteristic with respect to $l$-adic cohomology with compact
support, and $\mathcal{F}_{\chi _{triv}}$ denotes a suitable sheaf,
and $\mathbb{F}_{p^{e}}^{a}$ denotes an algebraic closure of
$\mathbb{F}_{p^{e}}$. Furthermore, they gave a description of the
poles of the multivariate local zeta functions in terms of the poles
of the topological zeta function: If $\boldsymbol{\rho}$ is a pole
of $\boldsymbol{Z}_{top}\left( \boldsymbol{s}\right)$, then for
almost all prime numbers $p$ there exist infinitely many unramified
extensions $\mathbb{K}_{e}$ of $\mathbb{Q}_{p}$ for which
$\boldsymbol{\rho}$ is a pole of $\boldsymbol{Z}\left(
\boldsymbol{s},\boldsymbol{f},\mathbb{K}_{e}\right)$, see [Theorem
(2.2) in \cite{denefandloeser}].
\subsection{String amplitudes and topological string zeta functions}

By using the fact that $\boldsymbol{Z}^{(N)}\left(
\underline{\boldsymbol{s}};I,0,\mathbb{K}_{e}\right)$ and
$\boldsymbol{Z}^{(N)}\left( \underline
{\boldsymbol{s}};T\smallsetminus I,1,\mathbb{K}_{e}\right)$ are
particular cases of $\boldsymbol{Z}\left(
\boldsymbol{s},\boldsymbol{f},\mathbb{K}_{e}\right)$, and by
applying (\ref{limir_p_goes_1}), we define
\begin{equation}
\boldsymbol{Z}_{top}^{(N)}\left(
\underline{\boldsymbol{s}};I,0\right) =\lim_{e\to
0}\boldsymbol{Z}^{(N)}\left(\underline{\boldsymbol{s}};I,0,\mathbb{K}_{e}\right)
\end{equation}
and
\begin{equation}
\boldsymbol{Z}_{top}^{(N)}\left(
\underline{\boldsymbol{s}};T\smallsetminus I,1\right) =\lim_{e\to
0}\boldsymbol{Z}^{(N)}\left(
\underline{\boldsymbol{s}};T\smallsetminus
I,1,\mathbb{K}_{e}\right),
\end{equation}
which are elements of $\mathbb{Q}\left(  s_{ij},i,j\in\left\{
1,\ldots,N-1\right\}  \right)  $, the field of rational functions in
the variables $s_{ij}$, $i,j \in \{1, \dots , N-1\}$ with
coefficients in $\mathbb{Q}$. Then, by using
(\ref{Formula_zeta_amplitude}) we define the {\it open string
N-point topological zeta function} as
\begin{equation}
\boldsymbol{Z}_{top}^{(N)}\left(  \underline{\boldsymbol{s}}\right)
=\sum_{I\subseteq T}\boldsymbol{Z}_{top}^{(N)}\left(  \underline
{\boldsymbol{s}};I,0\right) \boldsymbol{Z}_{top}^{(N)}\left(
\underline {\boldsymbol{s}};T\smallsetminus I,1\right).
\label{zeta_topologica}
\end{equation}
then we have the following result: the open string $N$-point
topological zeta function $\boldsymbol{Z}_{top}^{(N)}\left(
\underline{\boldsymbol{s}}\right)$ is a rational function of
$\mathbb{Q}\left( s_{ij},i,j\in\left\{ 1,\ldots ,N-1\right\}
\right)$ defined as (\ref{zeta_topologica}).

We define the {\it Denef-Loeser open string N-point amplitudes at
the tree level} as
\begin{equation}
\boldsymbol{A}_{top}^{(N)}\left(\underline{\boldsymbol{k}}\right)
=\boldsymbol{Z}_{top}^{(N)}\left(\underline{\boldsymbol{s}}\right)
\mid_{s_{ij}=\boldsymbol{k}_{i} \cdot \boldsymbol{k}_{j}},
\end{equation}
with $i\in\left\{ 1,\ldots,N-1\right\}$, $j\in T$ or $i,j\in T$,
where $T=\left\{  2,\ldots,N-2\right\}$. Thus the Koba-Nielsen
amplitudes are rational functions of the variables
$\boldsymbol{k}_{i} \cdot \boldsymbol{k}_{j}$, $i,j \in\left\{
1,\ldots,N\right\}$.

\subsection{Denef-Loeser open string four-point amplitudes}

In this subsection we calculate the open string $4$-point
topological zeta function. We recall that the open string $4$-point
zeta function is defined as
\begin{equation}
\boldsymbol{Z}^{(4)}(\underline{s},\mathbb{K}_{e})=
\int_{\mathbb{K}_{e}}|x_{2}|_{\mathbb{K}_{e}}^{s_{12}}|1-x_{2}|_{\mathbb{K}_{e}}^{s_{32}}dx_{2}.
\end{equation}
From (\ref{Formula_zeta_amplitude}) and (\ref{exponente}), we
calculate the contributions of each sector attached to $I\subseteq
T=\left\{ 2\right\}$:
$$
\boldsymbol{Z}^{(4)}(\underline{s},\mathbb{K}_{e}) =
\boldsymbol{Z}^{(4)}(\underline{s};\left\{
2\right\},0,\mathbb{K}_{e})\boldsymbol{Z}^{(4)}(\underline{s};\left\{
\varnothing \right\},1,\mathbb{K}_{e})
$$
\begin{equation}
+
p^{e(1+s_{12}+s_{32})}\boldsymbol{Z}^{(4)}(\underline{s};\left\{\varnothing
\right\},0,\mathbb{K}_{e})\boldsymbol{Z}^{(4)}(\underline{s};\left\{
2\right\},1,\mathbb{K}_{e}),
\end{equation}
where we recall that $\boldsymbol{Z}^{(4)}(\underline{s};\left\{
\varnothing \right\},0,\mathbb{K}_{e})=1,$
$\boldsymbol{Z}^{(4)}(\underline{s};\left\{ \varnothing
\right\},1,\mathbb{K}_{e})=1.$

By using the results given in Sections 3 and 4, we obtain
$$
\boldsymbol{Z}^{(4)}(\underline{s},\mathbb{K}_{e}) =
\boldsymbol{Z}^{(4)}(\underline{s};\left\{
2\right\},0,\mathbb{K}_{e}) +
p^{e(1+s_{12}+s_{32})}\boldsymbol{Z}^{(4)}(\underline{s};\left\{
2\right\},1,\mathbb{K}_{e})
$$
\begin{equation}
=
\int_{R_{\mathbb{K}_{e}}}|x_{2}|_{\mathbb{K}_{e}}^{s_{12}}|1-x_{2}|_{\mathbb{K}_{e}}^{s_{32}}dx_{2}
+
p^{e(1+s_{12}+s_{32})}\int_{R_{\mathbb{K}_{e}}}|x_{2}|_{\mathbb{K}_{e}}^{-2-s_{12}-s_{32}}dx_{2},
\end{equation}
where  $R_{\mathbb{K}_{e}}$ is the ring of integers of
$\mathbb{K}_{e}$ and
\begin{equation}
\boldsymbol{Z}^{(4)}(\underline{s};\left\{
2\right\},0,\mathbb{K}_{e})=1-2p^{-e}+ \frac{\left( 1-p^{-e}\right)
p^{e\left( -1-s_{12}\right) }}{1-p^{e\left( -1-s_{12}\right)
}}+\frac{\left( 1-p^{-e}\right) p^{e\left( -1-s_{32}\right)
}}{1-p^{e\left( -1-s_{32}\right) }}
\end{equation}
and
\begin{equation}
\boldsymbol{Z}^{(4)}(\underline{s};\left\{
2\right\},1,\mathbb{K}_{e}) =\frac{\left( 1-p^{-e}\right) p^{e\left(
1+s_{12}+s_{32}\right) }}{1-p^{e\left( 1+s_{12}+s_{32}\right) }}.
\end{equation}

Taking the limit $e$ approaches to zero, we obtain
\begin{equation}
\boldsymbol{Z}_{top}^{(4)}(\underline{s};\left\{ 2\right\},0)
=-1+\frac{1}{s_{12}+1}+\frac{1}{s_{32}+1}
\end{equation}
and
\begin{equation}
\boldsymbol{Z}_{top}^{(4)}(\underline{s};\left\{ 2\right\},1)
=-\frac{1}{s_{12}+s_{32}+1}.
\end{equation}
Consequently
\begin{equation}
\boldsymbol{Z}_{top}^{\left(
4\right)}(\underline{\boldsymbol{s}})=-1+\frac{
1}{s_{12}+1}+\frac{1}{s_{32}+1}-\frac{1}{s_{12}+s_{32}+1}.
\end{equation}
By using the kinematic relations
$\boldsymbol{k}_{1}+...+\boldsymbol{k}_{4}=0$ and
$\boldsymbol{k}_{i}^{2}=2$ we get $\boldsymbol{k}_{1} \cdot
\boldsymbol{k}_{2}+ \boldsymbol{k}_{3} \cdot \boldsymbol{k}_{2} +
1=-1-\boldsymbol{k}_{2} \cdot \boldsymbol{k}_{4},$ thus the
Denef-Loeser string $4$-point amplitude is given by
\begin{equation}
\boldsymbol{A}_{top}^{\left( 4\right) }(\underline{\boldsymbol{k}})
=Z_{top}^{\left( 4\right) }(\underline{\boldsymbol{k}})
=-1+\frac{1}{\boldsymbol{k}_{1} \cdot
\boldsymbol{k}_{2}+1}+\frac{1}{\boldsymbol{k}_{3} \cdot
\boldsymbol{k}_{2}+1}+\frac{1}{\boldsymbol{k}_{2} \cdot
\boldsymbol{k}_{4}+1}. \label{4amplitudes}
\end{equation}
This result is precisely the one that is obtained by finding the
scattering amplitudes from the resulting theory in the limit $p \to
1$ as we described in Section 2.

\subsection{Denef-Loeser open string five-point amplitudes}

Consider the case of Koba-Nielsen $5-$point amplitudes. The open
string $5$ -point zeta function is given by
\begin{equation}
\boldsymbol{Z}^{\left( 5\right)
}(\underline{\boldsymbol{s}},\mathbb{K}
_{e})=\int_{\mathbb{K}_{e}^{2}}|x_{2}|_{\mathbb{K}_{e}}^{s_{12}}|x_{3}|_{\mathbb{K}_{e}}^{s_{13}}
|1-x_{2}|_{\mathbb{K}_{e}}^{s_{42}}|1-x_{3}|_{\mathbb{K}_{e}}^{s_{43}}
|x_{2}-x_{3}|_{\mathbb{K}_{e}}^{s_{23}}dx_{2}dx_{3}.
\end{equation}
Formulae (\ref{Formula_zeta_amplitude})$-$(\ref{exponente}) require
an explicit description of the sectors attached to all the subsets
$I$ of $T=\{2,3\}$, i.e. for any $I \in
\{\{2\},\{3\},\{2,3\},\varnothing \}$. For instance, the sector
corresponding  to  $T=\{2,3\}$ is  $Sect(T)=\left\{ \left(
x_{2},x_{3}\right) \in \mathbb{K}_{e}^{2}; \
|x_2|_{\mathbb{K}_e}\leq 1 \ \rm{and} \ |x_3|_{\mathbb{K}_e}\leq
1\right\}.$ An explicit description of all the sectors is given in
Table 1.

The open string $5$-point topological zeta function is defined as
\begin{equation}
\boldsymbol{Z}_{top}^{(5)}\left( \underline{\boldsymbol{s}}\right)
=\sum_{I\subseteq T}\boldsymbol{Z}_{top}^{(5)}\left( \underline{
\boldsymbol{s}};I,0\right) \boldsymbol{Z}_{top}^{(5)}\left(
\underline{ \boldsymbol{s}} ;T\smallsetminus I,1\right).
\end{equation}
Table 2 contains explicit formulae for all the integrals
$\boldsymbol{Z}_{top}^{(5)}\left(\underline{\boldsymbol{s}}
;I,0\right)$ and $\boldsymbol{Z}_{top}^{(5)}\left(
\underline{\boldsymbol{s}};T\smallsetminus I,1\right)$.

\begin{table}[tbp]
\centering
\begin{tabular}{|c|c|c|}
\hline $I$ & $I^{c}$ & $Sect(I)$ \\ \hline $\left\{ 2\right\} $ &
$\left\{ 3\right\} $ & $R_{\mathbb{K}_{e}}\times
\mathbb{K}_{e}\backslash R_{\mathbb{K}_{e}}$ \\ \hline
$\left\{ 3\right\} $ & $\left\{ 2\right\} $ & $\mathbb{K}_{e}\backslash R_{%
\mathbb{K}_{e}}\times R_{\mathbb{K}_{e}}$ \\ \hline
$\left\{ 2,3\right\} $ & $\varnothing $ & $R_{\mathbb{K}_{e}}\times R_{%
\mathbb{K}_{e}}$ \\ \hline
$\varnothing$ & $\left\{ 2,3\right\} $ & $\mathbb{K}_{e}\backslash R_{%
\mathbb{K}_{e}}\times \mathbb{K}_{e}\backslash R_{\mathbb{K}_{e}}$
\\ \hline
\end{tabular}%
\caption{In the table we enumerate the different subsets $I$, their
complements $T\smallsetminus I$ and their associated region
$Sect(I)$.} \label{tab:ione}
\end{table}

\begin{table}[tbp]
\centering
\begin{tabular}{|c|c|c|}
\hline
$I$ & $\boldsymbol{Z}_{top}^{(5)}\left( \underline{\boldsymbol{s}}%
;I,0\right) $ & $\boldsymbol{Z}_{top}^{(5)}\left( \underline{\boldsymbol{s}};%
T\smallsetminus I,1\right) $ \\ \hline
$\left\{ 2\right\} $ & $-1+\frac{1}{1+s_{12}}+\frac{1}{1+s_{42}}$ & $-\frac{1%
}{1+s_{13}+s_{43}+s_{23}}$ \\ \hline &  &  \\ \hline
$\left\{ 3\right\} $ & $-1+\frac{1}{1+s_{13}}+\frac{1}{1+s_{43}}$ & $-\frac{1%
}{1+s_{12}+s_{42}+s_{23}}$ \\ \hline &  &  \\ \hline
$\left\{ 2,3\right\} $ & $%
\begin{array}{c}
\left[
\frac{1}{1+s_{12}}+\frac{1}{1+s_{13}}+\frac{1}{1+s_{23}}-1\right]
\frac{1}{2+s_{12}+s_{13}+s_{23}} \\
+\frac{1}{1+s_{12}}\left[ \frac{1}{1+s_{43}}-1\right] +\frac{1}{1+s_{13}}%
\left[ \frac{1}{1+s_{42}}-1\right] + \\
2-\frac{1}{1+s_{23}}-\frac{1}{1+s_{42}}-\frac{1}{1+s_{43}}+ \\
\frac{1}{2+s_{42}+s_{43}+s_{23}}\left[ \frac{1}{1+s_{42}}+\frac{1}{1+s_{43}}+%
\frac{1}{1+s_{23}}-1\right]%
\end{array}%
$ & 1 \\ \hline
$\left\{ \varnothing \right\} $ & 1 & $%
\begin{array}{c}
-\frac{1}{2+s_{52}+s_{53}+s_{23}}\times \\
\left[
\begin{array}{c}
\frac{1}{1+s_{12}+s_{42}+s_{23}}+\frac{1}{1+s_{13}+s_{43}+s_{23}} \\
+\frac{1}{1+s_{23}}-1%
\end{array}
\right]%
\end{array}
$ \\ \hline
\end{tabular}
\caption{The topological zeta functions
$\boldsymbol{Z}_{top}^{(5)}\left(\protect\underline{\boldsymbol{s}}
;I,0\right)$ and $\boldsymbol{Z}_{top}^{(5)}\left(
\protect\underline{\boldsymbol{s}};T\smallsetminus I,1\right)$ is
written for each subset $I$ and its complement $T\smallsetminus I$.}
\label{tab:itwo}
\end{table}

Thus, the Denef-Loeser open string $5$-point amplitude is given by
\begin{equation}
\begin{array}{l}
\boldsymbol{A}_{top}^{\left( 5\right) }(
\underline{\boldsymbol{k}})=\left[
\frac{ 1}{1+\boldsymbol{k}_{1} \cdot \boldsymbol{k}_{2}}+\frac{1}{1+%
\boldsymbol{k}_{4} \cdot \boldsymbol{k}_{2}}-1\right] \left[ -\frac{1}{1+%
\boldsymbol{k}_{3} \cdot \boldsymbol{k}_{5}}\right] +\left[ -\frac{1}{1+%
\boldsymbol{k}_{2} \cdot \boldsymbol{k}_{5}}\right] \left[ \frac{1}{1+%
\boldsymbol{k}_{1} \cdot\boldsymbol{k}_{3}}+\frac{
1}{1+\boldsymbol{k}_{4}
\cdot \boldsymbol{k}_{3}}-1\right] \\
\\
\text{ \ \ \ \ \ \ \ \ \ \ \ \ \ } + \left[
\frac{1}{1+\boldsymbol{k}_{1}
\cdot \boldsymbol{k}_{2}}+\frac{1}{1+\boldsymbol{k}_{1} \cdot \boldsymbol{k}%
_{3}}+\frac{1}{1+\boldsymbol{k}_{2} \cdot \boldsymbol{k}_{3}}-1\right] \frac{%
1}{1+\boldsymbol{k}_{4} \cdot \boldsymbol{k}_{5}} +\frac{1}{1+\boldsymbol{k}%
_{1} \cdot \boldsymbol{k}_{2}} \left[ \frac{1}{1+\boldsymbol{k}_{4}
\cdot
\boldsymbol{k}_{3}}-1\right] \\
\\
\text{ \ \ \ \ \ \ \ \ \ \ \ \ \ } + \frac{1}{1+\boldsymbol{k}_{1}
\cdot
\boldsymbol{k}_{3}}\left[ \frac{1}{1+\boldsymbol{k}_{4} \cdot \boldsymbol{k}%
_{2}}-1\right] +2- \frac{1}{1+\boldsymbol{k}_{2} \cdot \boldsymbol{k}_{3}}-%
\frac{1}{1+\boldsymbol{k}_{4} \cdot \boldsymbol{k}_{2}} -\frac{1}{1+%
\boldsymbol{k}_{4} \cdot \boldsymbol{k}_{3}} \\
\\
\text{ \ \ \ \ \ \ \ \ \ \ \ \ }+ \frac{1}{1+\boldsymbol{k}_{1}
\cdot
\boldsymbol{k}_{5}}\left[ \frac{1}{1+\boldsymbol{k}_{4} \cdot \boldsymbol{k}%
_{2}}+\frac{1}{1+ \boldsymbol{k}_{4} \cdot \boldsymbol{k}_{3}}+\frac{1}{1+%
\boldsymbol{k}_{2} \cdot \boldsymbol{k}_{3}}-1\right] \\
\\
\text{ \ \ \ \ \ \ \ \ \ \ \ \ \ } -\frac{1}{1+\boldsymbol{k}_{1}
\cdot
\boldsymbol{k}_{4}}\left[ \frac{1}{1+\boldsymbol{k}_{5} \cdot\boldsymbol{k}%
_{2}}+\frac{1}{1+\boldsymbol{k}_{3} \cdot \boldsymbol{k}_{5}} +\frac{1}{1+%
\boldsymbol{k}_{2} \cdot \boldsymbol{k}_{3}}-1\right].%
\end{array}
\label{5amplitudes}
\end{equation}

\section{Final Remarks}
\label{sec:final}

In this article we have considered the limit $p\to 1$ of $p$-adic
Koba-Nielsen amplitudes. In order to make mathematical sense of this
limit, we used the theory of topological zeta functions introduced
by Denef and Loeser. This requires to extend the $p$-adic
Koba-Nielsen amplitudes to unramified extensions of $\mathbb{Q}_p$,
more precisely to $\mathbb{K}_e$ the unique unramified extension of
degree $e$ of $\mathbb{Q}_p$, and then to express these amplitudes
as a finite sum of multivariate Igusa's zeta functions, see formulae
(\ref{Formula_zeta_amplitude})$-$(\ref{exponente}). This step is
carried out using the results of \cite{Bocardo-Gaspar:2016zwx},
however, we do not need the convergence of the $p$-adic Koba-Nielsen
amplitudes.

In this setting, using results due to Denef and Loeser, the limit $p
\to 1$ becomes the limit $e \to 0$. The computation of this last
limit requires explicit formulae for certain multivariate Igusa's
zeta functions, the required formulae were obtained using results
due to Denef. By taking the limit $e \to 0$ in the Koba-Nielsen
amplitudes over $\mathbb{K}_e$, we obtain the corresponding string
amplitudes. We computed explicitly the 4 and 5-point Denef-Loeser
string amplitudes, see formulae (\ref{4amplitudes}) and
(\ref{5amplitudes}). These amplitudes coincide exactly with the
quantum field derivation from the Gerasimov-Shatashvili Lagrangian
as presented in Section 2.

The topological zeta functions are particular cases of the motivic
Igusa's zeta functions constructed by Denef and Loeser in
\cite{motivicDL} using the theory of motivic integration, see
\cite{motivicintegrationDL}. Consequently, we can assert that there
exist motivic Koba-Nielsen amplitudes which specializes to the
topological Koba-Nielsen amplitudes introduced here and to the
classical $p$-adic Koba-Nielsen amplitudes, however, we do not know
if these motivic amplitudes have any physical meaning.

Such as it was mentioned in the introduction of
\cite{Bocardo-Gaspar:2016zwx}, see also \cite{WVeysWZG2017}, there
are deep connections between local zeta functions with string
amplitudes and quantum field theory amplitudes that still are not
fully understood.

Another relevant research direction is to explore the relation
between the topological amplitudes introduced here with the
amplitudes coming from the BSFT Lagrangian proposed by Witten in
\cite{Witten:1992qy,Witten:1992cr}. We expect a relation due to the
work of Gerasimov and Shatashvili \cite{Gerasimov:2000zp}. It would
be also interesting to study the incorporation of a $B$-field to the
string amplitudes such as was worked out in \cite{Ghoshal:2004ay}.
In this article the amplitudes are modified by a noncommutative
parameter satisfying the Moyal bracket. Finally it would be also
interesting to study the interplay between $p$-adic amplitudes in
field theory \cite{Zabrodin:1988ep}, AdS/CFT correspondence
\cite{Gubser:2016guj} and the renormalization group in discrete
world-sheet in the limit $p \to 1$, see \cite{Ghoshal:2006te}, using
the methods introduced here.


 \vspace{.5cm}
\centerline{\bf Acknowledgments} \vspace{.5cm}

It is a pleasure to thank D. Ghoshal for useful comments and
suggestions.   The work W. Z.-G. was partially supported CONACyT
grant 250845. H. G.-C. thanks the hospitality of DCI Universidad de
Guanajuato, where part of this research was developed during a
sabbatical leave.

\appendix

\section{Non-Archimedean local fields}
\label{sec:ENLE}

In these appendices, we review some basic ideas and results on
non-Archimedean and multivariate local zeta functions that we use
along this article.

We recall that the field of rational numbers $\mathbb{Q}$ admits two
types of norms: the Archimedean norm (the usual absolute value), and
the non-Archimedean norms (the $p$-adic norms) which are
parameterized by the prime numbers. The field of real numbers
$\mathbb{R}$ arises as the completion of $\mathbb{Q}$ with respect
to the Archimedean norm. Fix a prime number $p$, the $p$-adic norm
is defined as
\begin{equation}
\left\vert x\right\vert _{p}=\left\{
\begin{array}
[c]{lll}%
0 & \text{if} & x=0 \\
&  & \\
p^{-\gamma} & \text{if} & x=p^{\gamma}\frac{a}{b}\text{,}%
\end{array}
\right.
\end{equation}
where $a$ and $b$ are integers coprime with $p$. The integer
$\gamma:=ord(x)$, with $ord(0):=\infty$, is called the\textit{\
}$p$-\textit{adic order of} $x$. The field of $p$-adic numbers
$\mathbb{Q}_{p}$ is defined as the completion of the field of
rational numbers $\mathbb{Q}$ with respect to the $p$-adic norm
$|\cdot|_{p}.$

A non-Archimedean local field $\mathbb{K}$ is a locally compact
topological field with respect to a non-discrete topology, which
comes from a norm $\left\vert \cdot\right\vert_{\mathbb{K}}$
satisfying
\begin{equation}
\left\vert x+y\right\vert _{\mathbb{K}}\leq\max\left\{  \left\vert
x\right\vert _{\mathbb{K}},\left\vert y\right\vert
_{\mathbb{K}}\right\},
\end{equation}
for $x,y \in \mathbb{K}$. A such  norm is called an {\it ultranorm
or non-Archimedean}. Any non-Archimedean local field $\mathbb{K}$ of
characteristic zero is isomorphic (as a topological field) to a
finite extension of $\mathbb{Q}_{p}$, and it is called a $p$-adic
field. The field $\mathbb{Q}_{p}$ is the basic example of
non-Archimedean local field of characteristic zero. In the case of
positive characteristic, $\mathbb{K}$ is isomorphic to a finite
extension of the field of formal Laurent series
$\mathbb{F}_{q}\left(\left(T\right) \right)$ over a finite field
$\mathbb{F}_{q}$, where $q$ is a power of a prime number $p$.

In this article we work only with non-Archimedean fields
$\mathbb{K}$ of characteristic zero. Thus from now on $\mathbb{K}$
denotes one of these fields. The {\it ring of integers} of
$\mathbb{K}$ is defined as
\begin{equation}
R_{\mathbb{K}}=\left\{ x\in \mathbb{K};\left\vert
x\right\vert_{\mathbb{K}}\leq1\right\}.
\end{equation}

Geometrically $R_{\mathbb{K}}$ is the unit ball of the normed space
$\left( \mathbb{K},\left\vert \cdot\right\vert_{\mathbb{K}}\right)$.
This ring is a domain of principal ideals having a unique maximal
ideal, which is given by
\begin{equation}
P_{\mathbb{K}}=\left\{x\in \mathbb{K};\left\vert
x\right\vert_{\mathbb{K}}<1\right\}.
\end{equation}
We fix a generator $\pi$ of $P_{\mathbb{K}}$ i.e. $P_{\mathbb{K}}=
\pi R_{\mathbb{K}}$. A such generator is also called a {\it local
uniformizing parameter of} $\mathbb{K}$, and it plays the same role
as $p$ in $\mathbb{Q}_{p}.$

The {\it group of units} of $R_{\mathbb{K}}$ is defined as
\begin{equation}
R_{\mathbb{K}}^\times=\left\{  x\in R_{\mathbb{K}};\left\vert
x\right\vert_{\mathbb{K}}=1 \right\}.
\end{equation}
The natural map $R_{\mathbb{K}}\to
R_{\mathbb{K}}/P_{\mathbb{K}}\cong\mathbb{F}_{q}$ is called the {\it
reduction mod} $P_{\mathbb{K}}$. The quotient
$R_{\mathbb{K}}/P_{\mathbb{K}}\cong\mathbb{F}_{q}$, is a finite
field with $q=p^{f}$ elements, and it is called the {\it residue
field} of $\mathbb{K}$. Every non-zero element $x$ of $\mathbb{K}$
can be written uniquely as $x=\pi^{ord(x)}u$, $u\in
R_{\mathbb{K}}^{\times}$. We set $ord(0)=\infty$. The normalized
valuation of $\mathbb{K}$ is the mapping
\[
\begin{array}
[c]{ccc}%
\mathbb{K} & \to & \mathbb{Z}\cup\left\{  \infty\right\}  \\
x & \to & ord(x).
\end{array}
\]
Then $\left\vert x\right\vert_{\mathbb{K}}=q^{-ord(x)}$ and
$\left\vert \pi\right\vert_{\mathbb{K}}=q^{-1}$.

We fix $\mathfrak{S}\subset R_{\mathbb{K}}$ a set of representatives
of $\mathbb{F}_{q}$ in $R_{\mathbb{K}}$, i.e. $\mathfrak{S}$ is a
set which is mapped bijectively onto $\mathbb{F}_{q}$ by the
reduction $\operatorname{mod}$ $P_{\mathbb{K}}$. We assume that
$0\in\mathfrak{S}$. Any non-zero element $x$ of $\mathbb{K}$ can be
written as
\begin{equation}
x=\pi^{ord(x)}\sum\limits_{i=0}^{\infty}x_{i}\pi^{i},
\end{equation}
where $x_{i} \in\mathfrak{S}$ and $x_{0} \neq 0.$ This series
converges in the norm $\left\vert \cdot\right\vert_{\mathbb{K}}$.

We extend the norm $\left\vert \cdot\right\vert_{\mathbb{K}}$ to
$\mathbb{K}^{n}$ by taking
\begin{equation}
||\boldsymbol{x}||_{\mathbb{K}}:=\max_{1\leq i\leq
n}|x_{i}|_{\mathbb{K}},
\end{equation}
for $\boldsymbol{x}=(x_{1},\dots,x_{n})\in \mathbb{K}^{n}.$

We define $ord(\boldsymbol{x})=\min_{1\leq i\leq n}\{ord(x_{i})\}$,
then $||\boldsymbol{x}||_{\mathbb{K}}=q^{-ord(\boldsymbol{x})}$. The
metric space $\left(\mathbb{K}^{n},||\cdot||_{\mathbb{K}}\right)$ is
a complete ultrametric space.

For $r\in\mathbb{Z}$, denote by
$B_{r}^{n}(\boldsymbol{a})=\{\boldsymbol{x}\in
\mathbb{K}^{n};||\boldsymbol{x}-\boldsymbol{a}||_{\mathbb{K}}\leq
q^{r}\}$ \textit{the ball of radius }$q^{r}$ \textit{with center at}
$\boldsymbol{a}=(a_{1},\dots,a_{n})\in \mathbb{K}^{n}$, and take
$B_{r}^{n}(\boldsymbol{0}):=B_{r}^{n}$. Note that
$B_{r}^{n}(\boldsymbol{a})=B_{r}(a_{1})\times\cdots\times
B_{r}(a_{n})$, where $B_{r}(a_{i}):=\{x\in
\mathbb{K};|x_{i}-a_{i}|_{\mathbb{K}}\leq q^{r}\}$ is the
one-dimensional ball of radius $q^{r}$ with center at $a_{i}\in
\mathbb{K}$. The ball $B_{0}^{n}$ equals the product of $n$ copies
of $B_{0}=R_{\mathbb{K}}$. In addition,
$B_{r}^{n}(\boldsymbol{a})=\boldsymbol{a}+\left(
\pi^{-r}R_{\mathbb{K}}\right) ^{n}$. We also denote by
$S_{r}^{n}(\boldsymbol{a})=\{\boldsymbol{x}\in \mathbb{K}^{n}
;||\boldsymbol{x}-\boldsymbol{a}||_{\mathbb{K}}=q^{r}\}$ \textit{the
sphere of radius }$q^{r}$ \textit{with center at} $\boldsymbol{a}\in
\mathbb{K}^{n}$, and take
$S_{r}^{n}(\boldsymbol{0}):=S_{r}^{n}$. We notice that $S_{0}^{1}%
=R_{\mathbb{K}}^{\times}$ (the group of units of $R_{\mathbb{K}}$),
but $\left( R_{\mathbb{K}}^{\times}\right)^{n}\subsetneq S_{0}^{n}$,
for $n\geq2$. The balls and spheres are both open and closed subsets
in $\mathbb{K}^{n}$. In addition, two balls in $\mathbb{K}^{n}$ are
either disjoint or one is contained in the other.

The topological space
$\left(\mathbb{K}^{n},||\cdot||_{\mathbb{K}}\right)$ is totally
disconnected, i.e. the only connected subsets of $\mathbb{K}^{n}$
are the empty set and the points. A subset of $\mathbb{K}^{n}$ is
compact if and only if it is closed and bounded in $\mathbb{K}^{n}$.
The balls and spheres are compact subsets. Thus $\left(
\mathbb{K}^{n},||\cdot||_{\mathbb{K}}\right) $ is a locally compact
topological space.

As we mentioned before, any finite extension $\mathbb{K}$ of
$\mathbb{Q}_{p}$ is a non-Archimedean local field. Then
\begin{equation}
pR_{\mathbb{K}}=\pi^{m}R_{\mathbb{K}}, \ \ \ \ \ \ \ m\in
\mathbb{N}.
\end{equation}
If $m=1$ we say that $\mathbb{K}$ is a {\it unramified} extension of
$\mathbb{Q}_{p}.$ In other case, we say that $\mathbb{K}$ is a {\it
ramified} extension. It is well known that for every positive
integer $e$ there exists a unique unramified extension
$\mathbb{K}_e$ of $\mathbb{Q}_{p}$ of degree $e$, which means that
$\mathbb{K}_e$ is a $\mathbb{Q}_{p}$-vector space of dimension $e$.
From now on, $\pi$ denotes a local uniformizing parameter of
$\mathbb{K}_e$, thus $pR_{\mathbb{K}_{e}}=\pi R_{\mathbb{K}_{e}}$,
$R_{\mathbb{K}_{e}}/P_{\mathbb{K}_{e}} \cong\mathbb{F}_{p^{e}}$ and
$|\pi|_{\mathbb{K}_e} = p^{-e}$. For an in-depth exposition of
non-Archimedean local fields, the reader may consult
\cite{We,Taibleson}, see also \cite{Alberio et al,V-V-Z}.

\section{Multivariate Igusa zeta functions}

Let $\mathbb{K}$ be a $p$-adic field as before. Let
$f_{i}(\boldsymbol{x})\in \mathbb{K}\left[ x_{1},\ldots,x_{n}\right]
$ be a non-constant polynomial for $i=1,\ldots,r$, and \ let $\Phi$
be a Bruhat-Schwartz function, i.e. a locally constant
function with compact support. We set $\boldsymbol{f}=\left(  f_{1}%
,\ldots,f_{r}\right)  $ and $\boldsymbol{s}=\left(
s_{1},\ldots,s_{r}\right) \in\mathbb{C}^{r}$. The multivariate local
zeta function attached to $\left( f_{1},\ldots,f_{r},\Phi\right)$
(also called multivariate Igusa local zeta function) is defined as
\begin{equation}
\boldsymbol{Z}_{\Phi}\left(
\boldsymbol{s},\boldsymbol{f},\mathbb{K}\right)
=\int\limits_{\mathbb{K}^{n}\smallsetminus\cup_{i=1}^{r}f_{i}^{-1}(\boldsymbol{0})}
\Phi\left(  \boldsymbol{x}\right)  \prod\limits_{i=1}^{r}\left\vert
f_{i}(\boldsymbol{x})\right\vert_{\mathbb{K}}^{s_{i}}d^{n}\boldsymbol{x}
\end{equation}
for $\operatorname{Re}(s_{i})>0$, $i=1,\ldots,r$, here
$d^{n}\boldsymbol{x}$ denotes the normalized Haar measure of
$\mathbb{K}^{n}$. This integral defines a holomorphic function of
$\left( s_{1},\ldots,s_{r}\right)  $ in the half-space
$\operatorname{Re}(s_{i})>0$, $i=1,\ldots,r$. In the case $r=1$, the
local zeta functions were introduced by Weil, for general
$\boldsymbol{f}$ were first studied by Igusa \cite{Igusa}. In the
multivariate case, i.e. for $r\geq 1$, the local zeta functions were
studied by Loeser \cite{Loeser}. The Igusa local zeta functions are
related with the number of solutions of polynomial congruences
$\operatorname{mod}$ $p^{m}$ and with exponential sums
$\operatorname{mod}$ $p^{m}$. There are many intriguing conjectures
relating the poles of local zeta functions with the topology of
complex singularities, see e.g. \cite{Denef,Igusa}.

If $\Phi$ is the characteristic function of $R_{\mathbb{K}}^{n}$
we use the simplified notation $\boldsymbol{Z}\left(\boldsymbol{s}%
,\boldsymbol{f},\mathbb{K}\right)$.

\subsection{Embedded resolution of singularities}

In this section $\mathbb{L}$ is an arbitrary field of characteristic
zero and $f_{i}(\boldsymbol{x})\in
\mathbb{L}\left[\boldsymbol{x}\right]$, $\boldsymbol{x=} \left(
x_{1},\ldots,x_{n}\right)$, be a non-constant polynomial for
$i=1,\ldots,r$. The main tool in the study of local zeta functions
is Hironaka's resolution of singularities theorem \cite{Hi}. Put
$X=${\tt Spec} $\mathbb{L}\left[  \boldsymbol{x}\right]$ (the
$n$-dimensional affine space over
$\mathbb{L}$), $D=${\tt Spec} $\mathbb{L}\left[\boldsymbol{x}\right]  /\left(  \prod\nolimits_{i=1}%
^{r}f_{i}(\boldsymbol{x})\right)  $ (the divisor attached to
polynomials $f_{1}$,\ldots,$f_{r}$). An \textit{embedded resolution
of singularities} for $D$ over $\mathbb{L}$ consists of a pair
$(Y,h)$, where $Y$ is a smooth algebraic variety (an integral smooth
closed subscheme of the projective space over $X$), $h:Y\to X$ is
the natural map, which satisfies that the restriction
$h:Y\smallsetminus h^{-1}\left(  D\right)  \to X\smallsetminus D$ is
an isomorphism, and the reduced scheme $h^{-1}\left(
D\right)_{red\text{ }}$associated to $h^{-1}\left( D\right)  $ has
normal crossings, i.e. its irreducible components are smooth and
intersect transversally. Let $E_{i}$, $i\in T$, be the irreducible
components of $h^{-1}\left( D\right)  _{red\text{ }}$. For each
$i\in T$, let \ $N_{ij}$ be the multiplicity of $E_{i}$ in the
divisor $f_{j}\circ h$ on $Y$, and $v_{i}-1$ the multiplicity of
$E_{i}$ in the divisor $h^{\ast}\left( dx_{1}\wedge\ldots\wedge
dx_{n}\right)  $. The $\left( N_{i1},\ldots ,N_{ir},v_{i}\right)$,
$i\in T$, are called {\it the numerical data} of the resolution
$(Y,h)$. For $i\in T$ and $I\subset T$ we define
\begin{equation}
\overset{\circ}{E_{i}}=E_{i}\smallsetminus\bigcup\limits_{j\neq
i}E_{j}\text{,
\ \ }E_{I}=\bigcap\limits_{i\in I}E_{i}\text{, \ \ }\overset{\circ}{E_{I}%
}=E_{I}\smallsetminus\bigcup\limits_{j\in T\smallsetminus I}E_{j}.
\end{equation}
If $I=\emptyset$, we put $E_{\emptyset}=Y$.

\subsection{Rationality of local zeta functions}

\noindent {\bf Theorem A} (Loeser \cite{Loeser}) Let $\mathbb{K}$ be
a $p$-adic field. The local zeta function
$\boldsymbol{Z}_{\Phi}\left(
\boldsymbol{s},\boldsymbol{f},\mathbb{K}\right) $ admits \ a \
meromorphic \ continuation to \ the \ whole $\mathbb{C}^{r}$ as a
rational function of $q^{-s_{1}}, \ldots ,q^{-s_{r}}$, more
precisely
\begin{equation}
\boldsymbol{Z}_{\Phi}\left(
\boldsymbol{s},\boldsymbol{f},\mathbb{K}\right)
=\frac{P_{\Phi}\left( q^{-s_{1}},\ldots,q^{-s_{r}}\right)  }{\prod
\limits_{i\in T}\left( 1-q^{-v_{i}-\sum_{j=1}^{r}N_{ij}s_{j}}\right)
},
\end{equation}
where $P_{\Phi}$ is a polynomial in the variables
$q^{-s_{1}},\ldots,q^{-s_{r}}$. The real parts of the poles of
$\boldsymbol{Z}_{\Phi}\left(\boldsymbol{s},\boldsymbol{f},{\mathbb{K}}\right)$
belong to a union of hyperplanes of the form
\begin{equation}
v_{i}+\sum\limits_{j=1}^{r}N_{ij}\operatorname{Re}\left(
s_{j}\right) =0, \ \ \ \ \ i \in T.
\end{equation}

\vskip .5truecm

\noindent{\bf Theorem B} (Denef \cite{DL3})

Let $f_{i}(\boldsymbol{x})\in \mathbb{Z}\left[ \boldsymbol{x}\right]
$, $\boldsymbol{x=}\left( x_{1},\ldots ,x_{n}\right)$, be a
non-constant polynomial for $i=1,\ldots ,r$. Let $\left(Y,h\right)$
be an embedded resolution of singularities for $D=Spec$
$\mathbb{Q}\left[ \boldsymbol{x}\right] /\left(
\prod\nolimits_{i=1}^{r}f_{i}(\boldsymbol{x})\right)$ over
$\mathbb{Q}$, with numerical data $\left\{ \left( N_{i1},\ldots
,N_{ir},v_{i}\right) ;i\in T\right\}$. Then, there exists a finite
set of primes $S\subset \mathbb{Z}$ such that for any
non-Archimedean local field $\mathbb{K}\supset \mathbb{Q}$ with
$P_{\mathbb{K}}\cap \mathbb{Z} \notin S$, we have
\begin{equation}
\boldsymbol{Z}\left( \boldsymbol{s},\boldsymbol{f},\mathbb{K}\right)
=q^{-n}\sum\limits_{I\subseteq T}c_{I}\left( \mathbb{K}\right)
\prod\limits_{i\in I} \frac{\left( q-1\right)
q^{-v_{i}-\sum_{j=1}^{r}N_{ij}s_{j}}}{1-q^{-v_{i}-\sum_{j=1}^{r}N_{ij}s_{j}}},
\label{Explicit_for_1}
\end{equation}
where $q=q\left(\mathbb{K}\right)$ denotes the cardinality of the
residue field $\overline{\mathbb{K}}$ and
\begin{equation}
c_{I}(\mathbb{K})=Card\left\{ a\in \overline{Y}\left(
\overline{\mathbb{K}}\right) ;a\in \overline{E}_{i}\left(
\overline{\mathbb{K}}\right) \Leftrightarrow i\in I\text{ }
\right\}, \label{Explicit_for_2}
\end{equation}
where the bar denotes the reduction mod $P_{\mathbb{K}}$ for which
we refer to \cite[Sec. 2]{DL3}.

\newpage




\begin{thebibliography}{99}
\bibitem{Hlousek:1988vu}
  Z.~Hlousek and D.~Spector,
  ``$p$-adic string theory,''
  Annals Phys.\  {\bf 189}, 370 (1989).
  doi:10.1016/0003-4916(89)90170-X

\bibitem{Brekke:1993gf}
  L.~Brekke and P.~G.~O.~Freund,
  ``$p$-adic numbers in physics,''
  Phys.\ Rept.\  {\bf 233}, 1 (1993).
  doi:10.1016/0370-1573(93)90043-D

\bibitem {V-V-Z}
V. S. Vladimirov, V.I. Volovich, E.I. Zelenov, {\it p-adic analysis
and mathematical physics}. World Scientific, 1994.

\bibitem{Dragovich:2017kge}
  B.~Dragovich, A.~Y.~Khrennikov, S.~V.~Kozyrev, I.~V.~Volovich and E.~I.~Zelenov,
  ``$p$-Adic Mathematical Physics: The First 30 Years,''
  Anal.\ Appl.\  {\bf 9}, 87 (2017)
  doi:10.1134/S2070046617020017
  [arXiv:1705.04758 [math-ph]].


\bibitem{Gubser:2016guj}
  S.~S.~Gubser, J.~Knaute, S.~Parikh, A.~Samberg and P.~Witaszczyk,
  ``$p$-adic AdS/CFT,''
  Commun.\ Math.\ Phys.\  {\bf 352}, no. 3, 1019 (2017)
  doi:10.1007/s00220-016-2813-6
  [arXiv:1605.01061 [hep-th]].

\bibitem{Heydeman:2016ldy}
  M.~Heydeman, M.~Marcolli, I.~Saberi and B.~Stoica,
  ``Tensor networks, $p$-adic fields, and algebraic curves: arithmetic and the AdS$_3$/CFT$_2$ correspondence,''
  arXiv:1605.07639 [hep-th].

\bibitem{Gubser:2016htz}
  S.~S.~Gubser, M.~Heydeman, C.~Jepsen, M.~Marcolli, S.~Parikh, I.~Saberi, B.~Stoica and B.~Trundy,
  ``Edge length dynamics on graphs with applications to $p$-adic AdS/CFT,''
  JHEP {\bf 1706}, 157 (2017)
  doi:10.1007/JHEP06(2017)157
  [arXiv:1612.09580 [hep-th]].

\bibitem{Dutta:2017bja}
  P.~Dutta, D.~Ghoshal and A.~Lala,
  ``On the Exchange Interactions in Holographic $p$-adic CFT,''
  arXiv:1705.05678 [hep-th].

\bibitem{Sen:1998sm}
  A.~Sen,
  ``Tachyon condensation on the brane anti-brane system,''
  JHEP {\bf 9808}, 012 (1998)
  doi:10.1088/1126-6708/1998/08/012
  [hep-th/9805170].

\bibitem{Sen:1999xm}
  A.~Sen,
  ``Universality of the tachyon potential,''
  JHEP {\bf 9912}, 027 (1999)
  doi:10.1088/1126-6708/1999/12/027
  [hep-th/9911116].

\bibitem{Sen:1999nx}
  A.~Sen and B.~Zwiebach,
  ``Tachyon condensation in string field theory,''
  JHEP {\bf 0003}, 002 (2000)
  doi:10.1088/1126-6708/2000/03/002
  [hep-th/9912249].

\bibitem{Berkovits:2000hf}
  N.~Berkovits, A.~Sen and B.~Zwiebach,
  ``Tachyon condensation in superstring field theory,''
  Nucl.\ Phys.\ B {\bf 587}, 147 (2000)
  doi:10.1016/S0550-3213(00)00501-0
  [hep-th/0002211].

\bibitem{Ghoshal:2000dd}
  D.~Ghoshal and A.~Sen,
  ``Tachyon condensation and brane descent relations in $p$-adic string theory,''
  Nucl.\ Phys.\ B {\bf 584}, 300 (2000)
  doi:10.1016/S0550-3213(00)00377-1
  [hep-th/0003278].

\bibitem{Berera:1992tm}
  A.~Berera,
  ``Unitary string amplitudes,''
  Nucl.\ Phys.\ B {\bf 411}, 157 (1994).
  doi:10.1016/0550-3213(94)90057-4

\bibitem{Witten:2013pra}
  E.~Witten,
  ``The Feynman $i \varepsilon$ in String Theory,''
  JHEP {\bf 1504}, 055 (2015)
  doi:10.1007/JHEP04(2015)055
  [arXiv:1307.5124 [hep-th]].

\bibitem{Brekke:1988dg}
  L.~Brekke, P.~G.~O.~Freund, M.~Olson and E.~Witten,
``Nonarchimedean String Dynamics,''
  Nucl.\ Phys.\ B {\bf 302}, 365 (1988).
  doi:10.1016/0550-3213(88)90207-6

\bibitem{Frampton:1987sp}
  P.~H.~Frampton and Y.~Okada,
  ``The $p$-adic String $N$-Point Function,''
  Phys.\ Rev.\ Lett.\  {\bf 60}, 484 (1988).
  doi:10.1103/PhysRevLett.60.484

\bibitem{Moeller:2002vx}
  N.~Moeller and B.~Zwiebach,
  ``Dynamics with infinitely many time derivatives and rolling tachyons,''
  JHEP {\bf 0210}, 034 (2002)
  doi:10.1088/1126-6708/2002/10/034
  [hep-th/0207107].

\bibitem{Barnaby:2006hi}
  N.~Barnaby, T.~Biswas and J.~M.~Cline,
  ``$p$-adic Inflation,''
  JHEP {\bf 0704}, 056 (2007)
  doi:10.1088/1126-6708/2007/04/056
  [hep-th/0612230].

\bibitem{Freund:1987ck}
  P.~G.~O.~Freund and E.~Witten,
  ``Adelic String Amplitudes,''
  Phys.\ Lett.\ B {\bf 199}, 191 (1987).
  doi:10.1016/0370-2693(87)91357-8

\bibitem{Gerasimov:2000zp}
  A.~A.~Gerasimov and S.~L.~Shatashvili,
  ``On exact tachyon potential in open string field theory,''
  JHEP {\bf 0010}, 034 (2000)
  doi:10.1088/1126-6708/2000/10/034
  [hep-th/0009103].

  \bibitem{Spokoiny:1988zk}
  B.~L.~Spokoiny,
  ``Quantum Geometry of Nonarchimedean Particles and Strings,''
  Phys.\ Lett.\ B {\bf 208}, 401 (1988).
  doi:10.1016/0370-2693(88)90637-5

\bibitem{Witten:1992qy}
  E.~Witten,
  ``On background independent open string field theory,''
  Phys.\ Rev.\ D {\bf 46}, 5467 (1992)
  doi:10.1103/PhysRevD.46.5467
  [hep-th/9208027].

\bibitem{Witten:1992cr}
  E.~Witten,
  ``Some computations in background independent off-shell string theory,''
  Phys.\ Rev.\ D {\bf 47}, 3405 (1993)
  doi:10.1103/PhysRevD.47.3405
  [hep-th/9210065].

\bibitem{Minahan:2000ff}
  J.~A.~Minahan and B.~Zwiebach,
  ``Field theory models for tachyon and gauge field string dynamics,''
  JHEP {\bf 0009}, 029 (2000)
  doi:10.1088/1126-6708/2000/09/029
  [hep-th/0008231].

\bibitem{Ghoshal:2004dd}
  D.~Ghoshal,
  ``Exact noncommutative solitons in $p$-adic strings and BSFT,''
  JHEP {\bf 0409}, 041 (2004)
  doi:10.1088/1126-6708/2004/09/041
  [hep-th/0406259].

\bibitem{Ghoshal:2006te}
  D.~Ghoshal,
  ``$p$-adic string theories provide lattice discretization to the ordinary string worldsheet,''
  Phys.\ Rev.\ Lett.\  {\bf 97}, 151601 (2006)
  doi:10.1103/PhysRevLett.97.151601
  [hep-th/0606082].

\bibitem{Bocardo-Gaspar:2016zwx}
  M.~Bocardo-Gaspar, H.~Garc\'{\i}a-Compe\'an and W.~A.~Z\'u\~niga-Galindo,
  ``Regularization of $p$-adic String Amplitudes, and Multivariate Local Zeta Functions,''
  arXiv:1611.03807 [math-ph].

\bibitem{Igusa}
J.-I. Igusa, {\it An introduction to the theory of local zeta
functions}, AMS/IP Studies in Advanced Mathematics, 2000.

\bibitem{Meuser}
D. Meuser, {\it  A survey of Igusa's local zeta function}, Amer. J.
Math. (2016), no. 1, 149179.

\bibitem{denefandloeser}
J. Denef and  F. Loeser, {\it Caract\'{e}ristiques
D'Euler-Poincar\'{e}, Fonctions Zeta locales et modifications
analytiques}, (1998), no. 3, 505-537.

\bibitem{trossmann}
T. Rossmann, ``Computing topological zeta functions of groups,
algebras, and modules, I,'' Proc. Lond. Math. Soc. (3) {\bf 110}
(2015), no. 5, 1099-1134.

\bibitem{Fuchs:2008cc}
  E.~Fuchs and M.~Kroyter,
  ``Analytical Solutions of Open String Field Theory,''
  Phys.\ Rept.\  {\bf 502}, 89 (2011)
  doi:10.1016/j.physrep.2011.01.003
  [arXiv:0807.4722 [hep-th]].

\bibitem{Peskin:1995ev}
  M.~E.~Peskin and D.~V.~Schroeder,
  `An Introduction to quantum field theory,'' Addison-Wesley
  Publishing Company (1995).

\bibitem{motivicDL}
J. Denef and F. Loeser, ``Motivic Igusa zeta functions,'' J.
Algebraic Geom. {\bf 7} (1998), no. 3, 505-537.

\bibitem{motivicintegrationDL}
J. Denef and F. Loeser, ``Germs of arcs in singular algebraic
varieties and motivic integration,'' Invent. Math. {\bf 135} (1999),
no. 1, 201-232.

\bibitem{WVeysWZG2017}
W.~Veys and W.~A.~Z\'u\~niga-Galindo, ``Zeta functions and
oscillating integrals for meromorphic functions,'' Adv. Math. {\bf
311} (2017) 295-337.

\bibitem{Ghoshal:2004ay}
  D.~Ghoshal and T.~Kawano,
  ``Towards $p$-adic string in constant $B$-field,''
  Nucl.\ Phys.\ B {\bf 710}, 577 (2005)
  doi:10.1016/j.nuclphysb.2004.12.025
  [hep-th/0409311].

\bibitem{Zabrodin:1988ep}
  A.~V.~Zabrodin,
  ``Nonarchimedean Strings and Bruhat-tits Trees,''
  Commun.\ Math.\ Phys.\  {\bf 123}, 463 (1989).
  doi:10.1007/BF01238811

\bibitem{We} A. Weil, {\it Basic number theory}, Reprint of the second
(1973) edition. Classics in Mathematics. Springer-Verlag, Berlin,
1995.

\bibitem{Taibleson}
M.H. Taibleson, {\it Fourier analysis on local fields}, Princeton
University Press, 1975.

\bibitem{Alberio et al}
S. Albeverio, A. Yu. Khrennikov, V.M. Shelkovich, {\it Theory of
$p$-adic distributions linear and nonlinear models}.  London
Mathematical Society Lecture Note Series, 370. Cambridge University
Press, Cambridge, 2010.

\bibitem{Loeser} F. Loeser, {\it Fonctions z\^{e}ta locales d'Igusa \`{a} plusieurs
variables, int\'{e}gration dans les fibres, et discriminants}, Ann.
Sci. \'{E}cole Norm. Sup. (4) {\bf 22} (1989), no. 3, 435--471.

\bibitem{Denef} J. Denef, {\it Report on Igusa's Local Zeta Function}, S\'{e}minaire
Bourbaki 43 (1990-1991), exp. 741; Ast\'{e}risque 201-202-203
(1991), 359-386. Available at
http://www.wis.kuleuven.ac.be/algebra/denef.html.

\bibitem{Hi} H. Hironaka, {\it Resolution of singularities of an algebraic variety
over a field of characteristic zero}, Ann. Math., {\bf 79} (1969),
109--326.

\bibitem {DL3} J. Denef, On the degree of Igusa's local zeta function, Amer.
J. Math. {\bf 109} (1987), 991--1008.

\end{thebibliography}
\end{document}